\documentclass[referee]{aa}
\usepackage{graphics}
\usepackage{epsf}

\def\h{$h^{-1}$ Mpc}

\begin{document}

%\thesaurus{2(12.03.3; 12.03.4)}

\title{Nature and Environment of Very Luminous Galaxies}

\author{A. Cappi\inst{1} \and  C. Benoist\inst{2} \and  
L.N. da Costa\inst{3} \and S. Maurogordato\inst{4}}
\institute{
INAF, Osservatorio Astronomico di Bologna, via Ranzani 1, I--40127,
   Bologna, Italy \\
\email{cappi@bo.astro.it}
\and
CERGA, Observatoire de la C\^ote d'Azur, BP4229, Le Mont--Gros, 06304 Nice 
Cedex 4, France \\
\email{benoist@obs-nice.fr}
\and
European Southern Observatory, Karl Schwartzschildsrtra$\beta$e 2, D--85748 
Garching bei M\"unchen, Germany \\
\email{ldacosta@eso.org}
\and
CERGA, CNRS, Observatoire de la C\^ote d'Azur, BP4229, Le Mont--Gros, 06304 
Nice Cedex 4, France \\
\email{maurogor@obs-nice.fr}
}
\offprints{Alberto Cappi}
\date{Received ~ / Accepted ~}
\authorrunning{Cappi et al.}
\titlerunning{Nature and Environment of VLGs}

\abstract{
   The most luminous galaxies in the blue passband have a larger
   correlation amplitude than $L^*$ galaxies. They do not
   appear to be preferentially located in rich clusters or groups, 
   but a significant fraction of them  seem to be in systems which 
   include fainter members.
   We present an analysis of fields centered on 18 Very 
   Luminous Galaxies ($M_B \le -21$) selected from the Southern Sky Redshift 
   Survey 2, based on new observations and public data of the 2dF Galaxy 
   Redshift Survey; we present also additional data on a CfA VLG and on 
   Arp 127.
   We find that all the selected VLGs are physically associated to fainter 
   companions. Moreover, there is a relation between the VLG morphology
   (early or late) and the dynamical properties of the system, which reflects
   the morphology--density relation. 6 out of the 18 SSRS2
   VLGs are early type galaxies: 2 are in the center of rich Abell clusters 
   with velocity dispersion $\sigma \sim 600$ km/s, 
   and the other 4 are in poor clusters or groups with $\sigma \sim 300$.
   The VLG extracted from the CfA catalog is also an elliptical in a 
   Zwicky cluster. 
   The remaining 2/3 of the sample are late--type VLGs,
   generally found in poorer systems with a larger spread
   in velocity dispersion, from $\sim 100$ up to $\sim 750$ kms/s. 
   The low velocity dispersion, late--type VLG dominated systems 
   appear to be the analogous of our own Local Group.
   The possibile association of VLG systems to dark matter halos with mass 
   comparable to rich groups or clusters, as suggested by the comparable
   correlation amplitude, would imply significant differences in the 
   galaxy formation process. 
   This work also shows that observing fields around VLGs represents an 
   effective way of identifying galaxy systems which are not selected through
   other traditional techniques. 
}

\maketitle

%
% Section 1
%

\section{Introduction}

From the analysis of the Southern Sky Redshift Survey 2 
(SSRS2, da Costa et al. 1994), we have found that the amplitude 
of galaxy clustering increases significantly as a function of 
galaxy luminosity, 
but only when $L > L^*$ (Benoist et al. 1996; 
see also Willmer et al. 1998); moreover, the analysis of
high--order moments shows that the bias is not linear, without a 
significant dependence on scale (Benoist et al. 1999), 
analogously to the bias between clusters and galaxies 
(Cappi \& Maurogordato 1995).
We have also found that the clustering amplitude of the most 
luminous galaxies in the sample, 
having absolute magnitude $M_B \le -21$ (i.e. $L \geq 4 L^*$), which
we defined as Very Luminous Galaxies (VLGs),
is similar to that of clusters, with 
a correlation length $r_0 \sim 16$ \h.
From the analysis of the 2dF Galaxy Redshift Survey (2dFGRS)
Norberg et al. (2001) have confirmed the reality of luminosity segregation 
for galaxies more luminous than $L^*$, even if their most luminous galaxies 
show a correlation amplitude not so high as the SSRS2 VLGs.

The large value of the VLG correlation function could be explained
if VLGs were in clusters, as originally suggested by Hamilton (1984): in this
case, most of them should be luminous ellipticals. However, in our statistical
analysis of the SSRS2 VLGs (Cappi et al. 1998, Paper I)
we have shown that the fractions of the different morphological types 
are comparable to lower luminosity samples, and that most of the VLGs are not 
in rich clusters. Only a minority of VLGs were found in known groups, but 
in most cases our visual inspection of the Digitized Sky Survey 
images revealed the presence of fainter companions, often with
signs of interaction.

In fact, if VLGs are neither in clusters nor in rich groups,
they can nevertheless be associated to poorer systems.
In biased galaxy formation (Kaiser 1984; Bardeen et al. 1986)
more massive halos, which represent
rare fluctuations of the matter density field, have a larger correlation 
amplitude. This suggests an association of VLGs to high density peaks;
the fact that they are dominant galaxies in poor systems with much fainter
galaxies can have interesting implications, concerning for example the 
efficiency of  galaxy formation or the overabundance of predicted subhalos 
relative to the observed dwarf satellites of the Milky Way and M31 
(see e.g. Moore et al. 1999): for this reason, it would be useful to have a 
statistical sample of galaxy systems similar to our own Local Group.

Unfortunately the properties of poor galaxy systems, 
especially those comparable to the Local Group, are not well known.
Zabludoff \& Mulchaey (2000) have
studied a sample of six nearby poor groups, finding evidence for a different 
luminosity function, with an increasing dwarf to giant galaxy ratio with the 
mass density of the system.
More generally, Zabludoff \& Mulchaey (1998, 2000) 
have examined the properties of 12 poor groups of galaxies with PSPC images: 
9 of them have diffuse X--ray emission and a bright, central elliptical galaxy.

Our approach is complementary, as it selects another class of poor 
systems. In fact, a main problem is the bias on the properties of the 
selected galaxy groups due to the adopted selection criteria.
A selection algorithm defines {\em a priori} the properties of the systems:
well known examples are Abell clusters
(Abell 1958) and Hickson compact groups (Hickson 1982).
The introduction of automated methods can give a more objective selection, 
but it is obviously impossible to recover a system if only its most 
luminous member is present in the original photometric catalogue. Moreover,
the typical friends--of--friends algorithms used to select 
``galaxy groups" require the detection of at least three neighbouring 
galaxies above the limiting magnitude of the catalogue 
(see e.g. White et al. 1999): this means that many groups are classified as
``binaries''. It is clear that the problem becomes critical when going towards
 poorer and more distant systems, with increasing contamination and spurious 
detections.

Let us take the best example of
VLG system, i.e. our Local Group, which contains two VLGs: M31 and 
our Milky Way (with respectively $M_V=-21.2$ and $M_V=-20.9$, 
see van den Bergh 1999 and references therein), and
is usually considered as a ``typical group''.
If this is true, it is surely not reflected in group
catalogs, for a simple reason: VLGs are rare galaxies. For example,
the third brightest galaxy of the Local Group M33 (with an absolute magnitude
$M_V=-18.9$) would become fainter than $m_V=15.5$ (i.e. fainter than
the limit of the SSRS2) already at $z \sim 0.025$. 
Therefore we do not know if the properties of 
our Local Group are really ``typical'', and only
looking deeper at VLG fields we can expect to find other groups 
similar to our own.

In order to construct a statistical sample including, among others, 
also systems comparable to our Local Group, we have decided to investigate 
the environment of the VLGs listed in our catalogue defined in paper I, 
which represent a volume--limited sample.
The first step is to measure redshifts of fainter galaxies in the
field of VLGs and determine the membership and velocity dispersion of the
systems: in this paper we discuss data concerning the fields of 
19 VLGs.
In section 2 we define our sample, with the VLG
fields we observed at the Observatoire de Haute Provence and the 
selection of galaxies around VLGs extracted from the 2dFGRS; 
section 3 presents the individual properties of the systems,
while in section 4 we discuss the relation between 
the VLGs and their environment. Our conclusions are in section 5.
In appendix A we also discuss Arp 127, a pair identified in our preliminary 
selection as possibly including a VLG, and in appendix B we give positions,
magnitudes and redshifts of 2dFGRS galaxies in our VLG systems. 

\section{Definition of the sample}

Three VLGs were selected from our catalogue of VLGs (see table 1 in paper I), 
while a fourth galaxy satisfying the VLG definition was
selected from the CfA catalogue (see Geller \& Huchra 1989); 
the 4 VLG fields were observed at OHP.
Other 15 VLG fields were extracted from the 2dFGRS public catalogue, 
which has an overlap with the SSRS2. While partially imposed by
observational constraints (see below), our selection
is random with respect to the VLG properties, and should be 
representative of the whole sample. In fact,
one third of the selected VLGs are early--type galaxies,
a fraction consistent with that of the total sample
published in paper I.

\begin{figure}
\caption[]{Finding charts of the VLG fields observed at OHP 
(the scale is approximately $30 \times 30$ square arcmin).}
\label{fig:ohpfields}
\centerline{VLG061: finding chart. \hskip 6.6cm VLG068: finding chart} 
{\epsfxsize=8.0cm \epsfbox{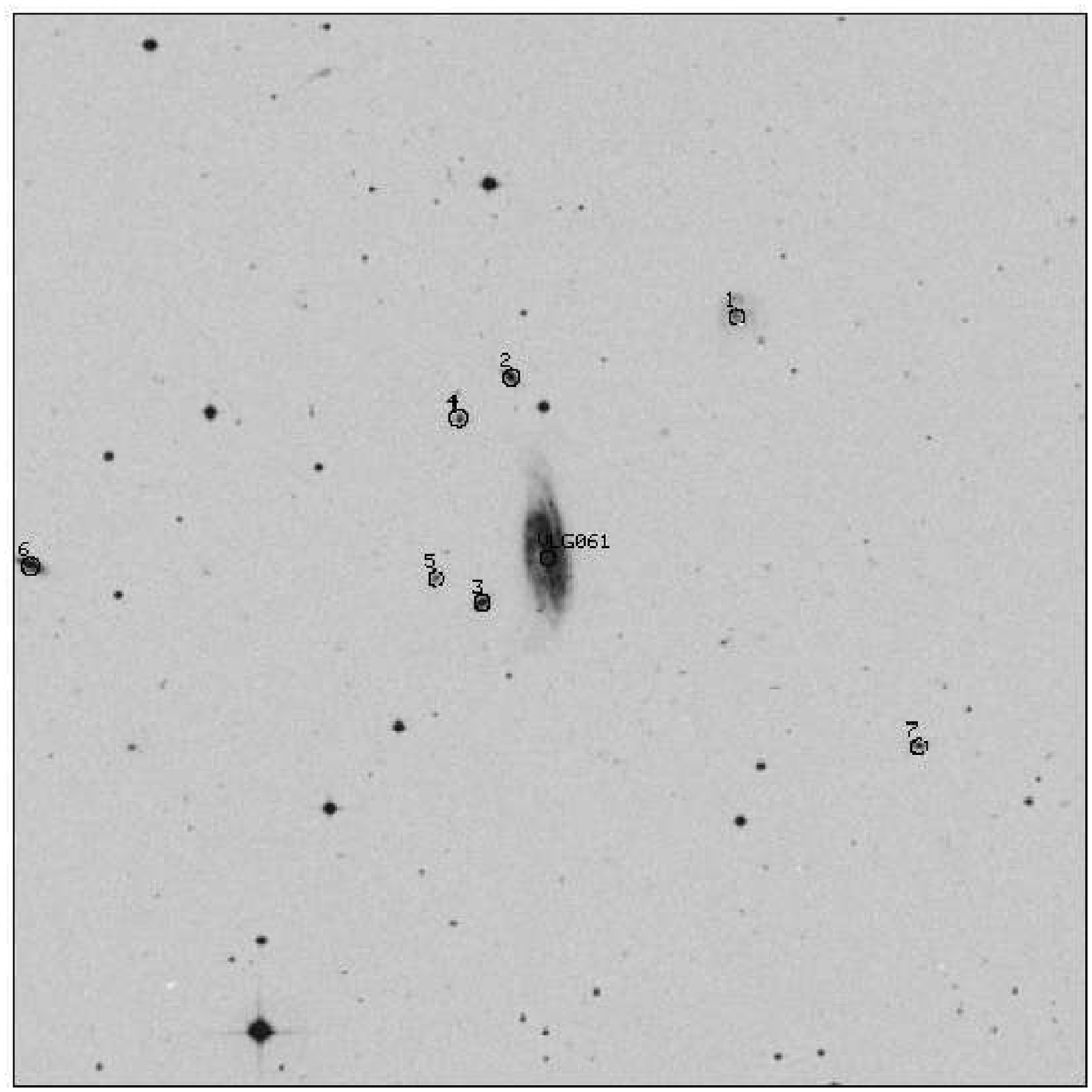} \hskip 1cm
\epsfxsize=8.0cm \epsfbox{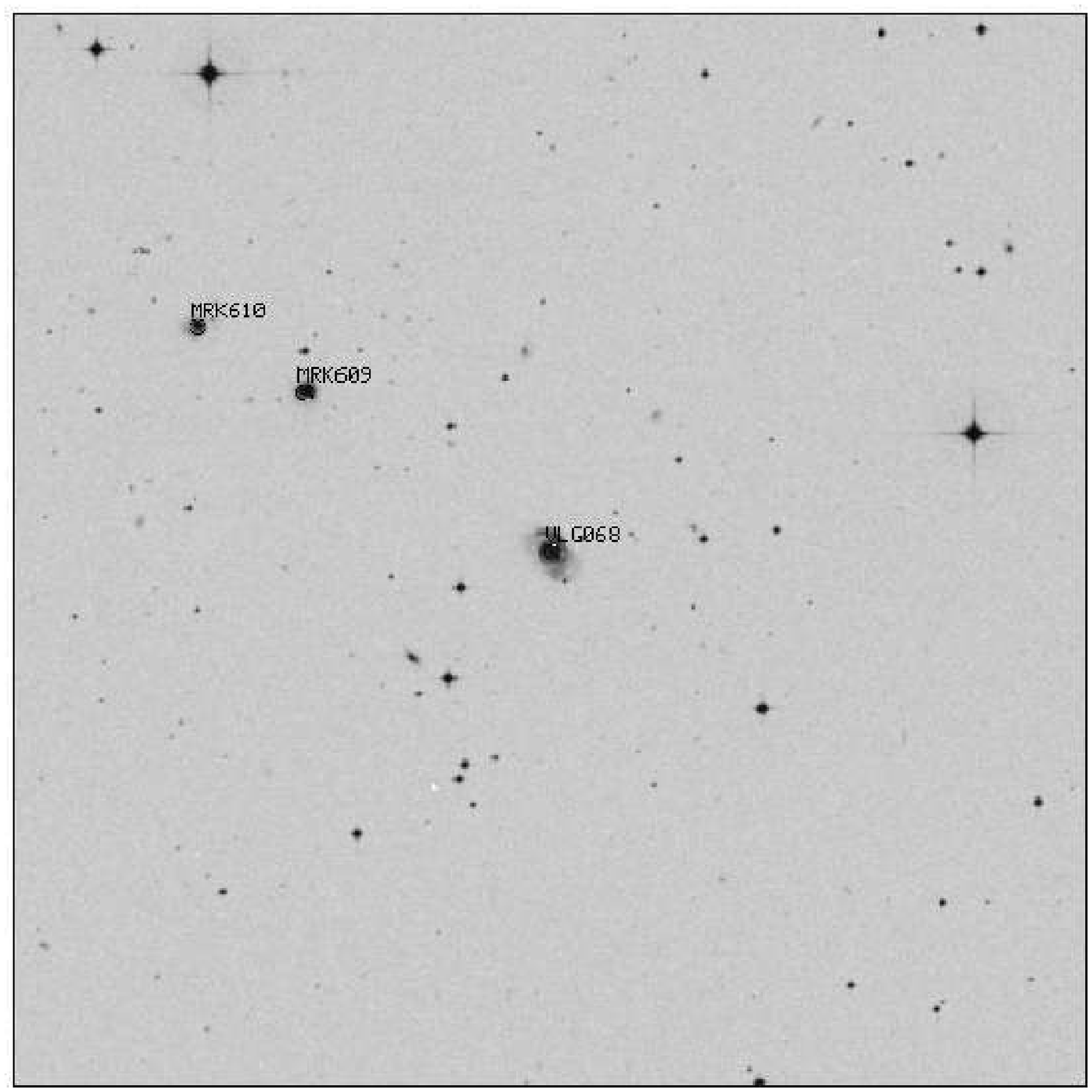}}
\centerline{VLG074: finding chart. \hskip 6.6cm Z1261: finding chart} 
{\epsfxsize=8.0cm \epsfbox{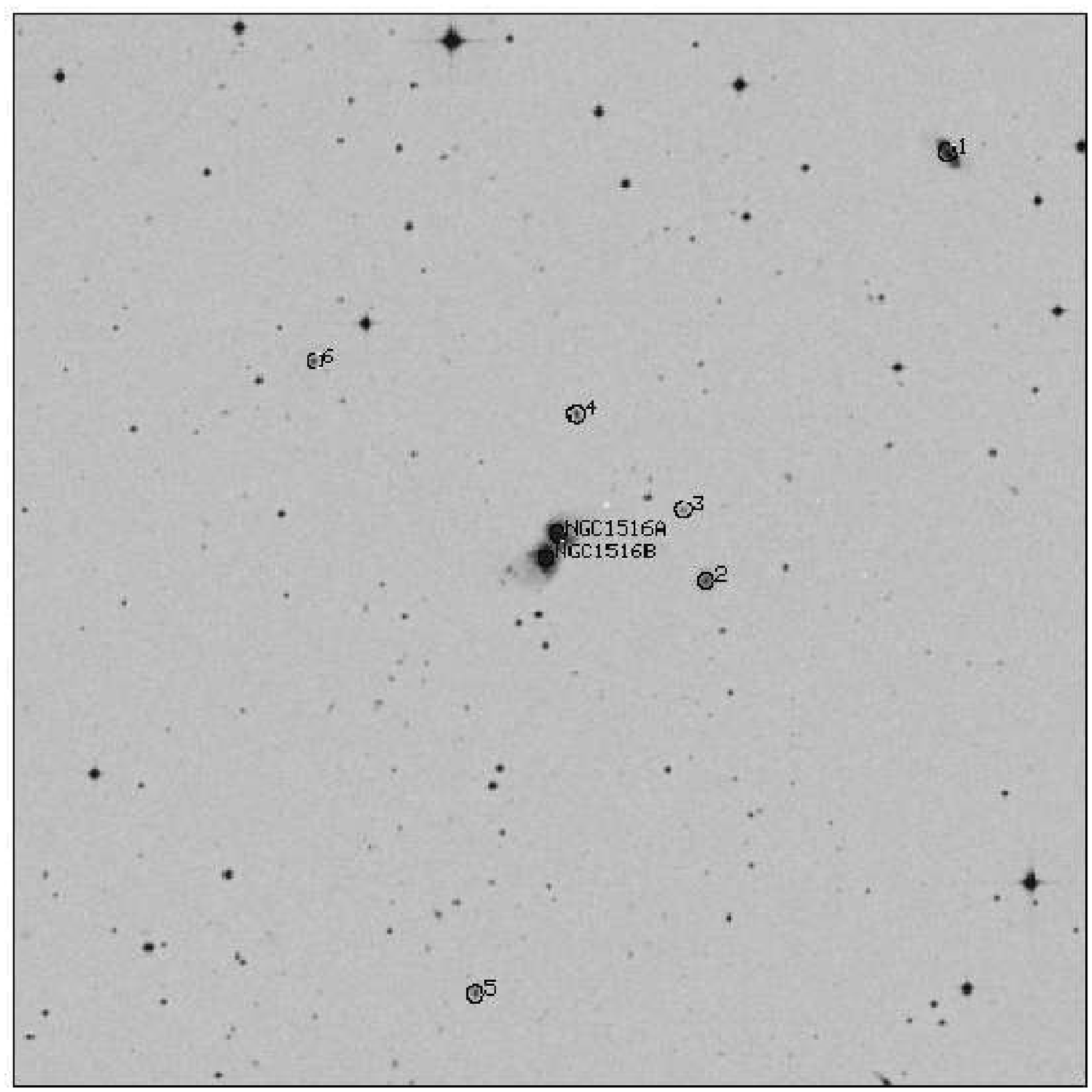} \hskip 1cm
\epsfxsize=8.0cm \epsfbox{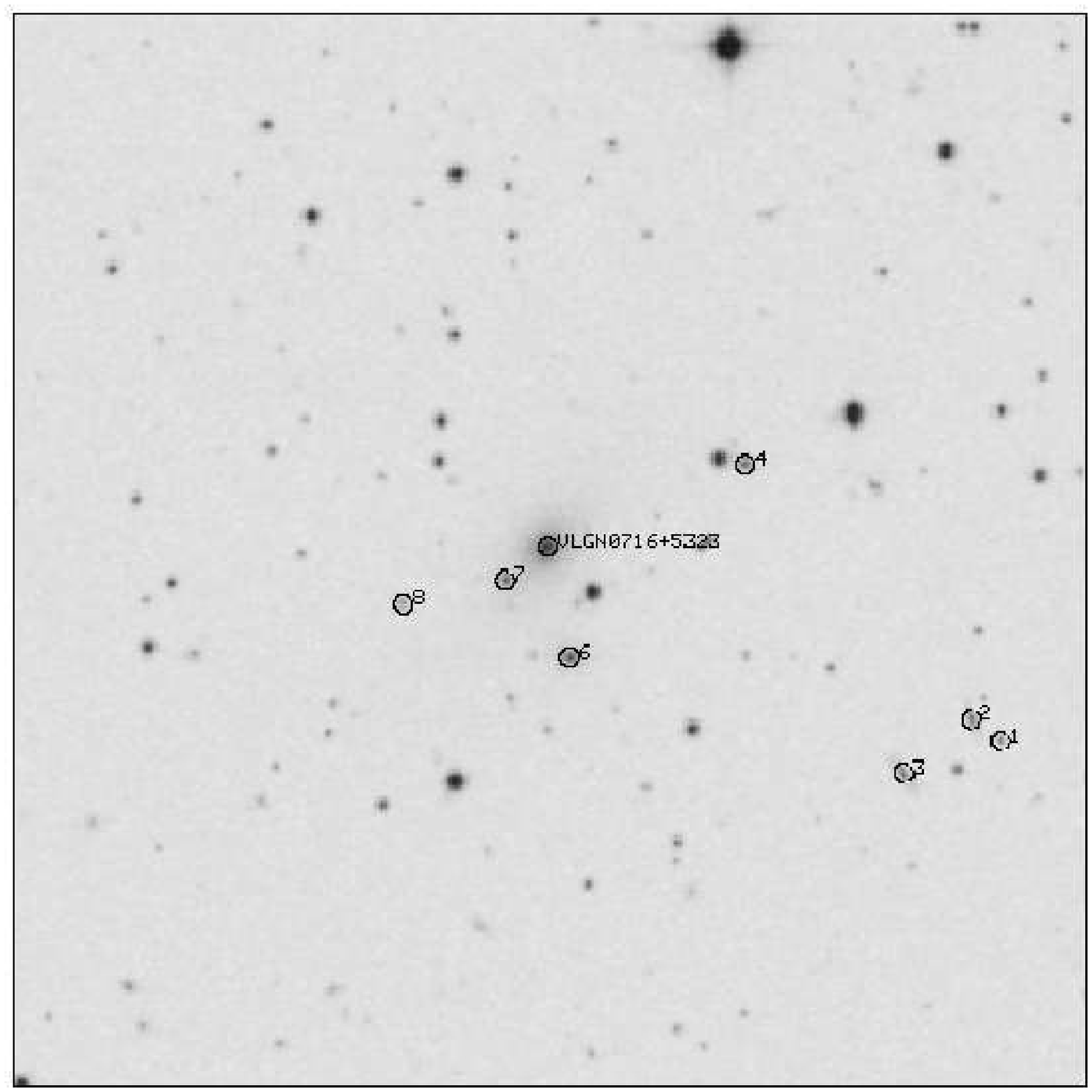}}
\end{figure}
\begin{figure}
\vskip 2cm
\label{fig:fields}
\caption[]{Finding charts of the VLG fields selected from the 2dFGRS
(the scale is approximately $40 \times 30$ square arcmin). \\}
\centerline{VLG014 \hskip 6.6cm VLG022}
  {\epsfxsize=8.0cm \epsfbox{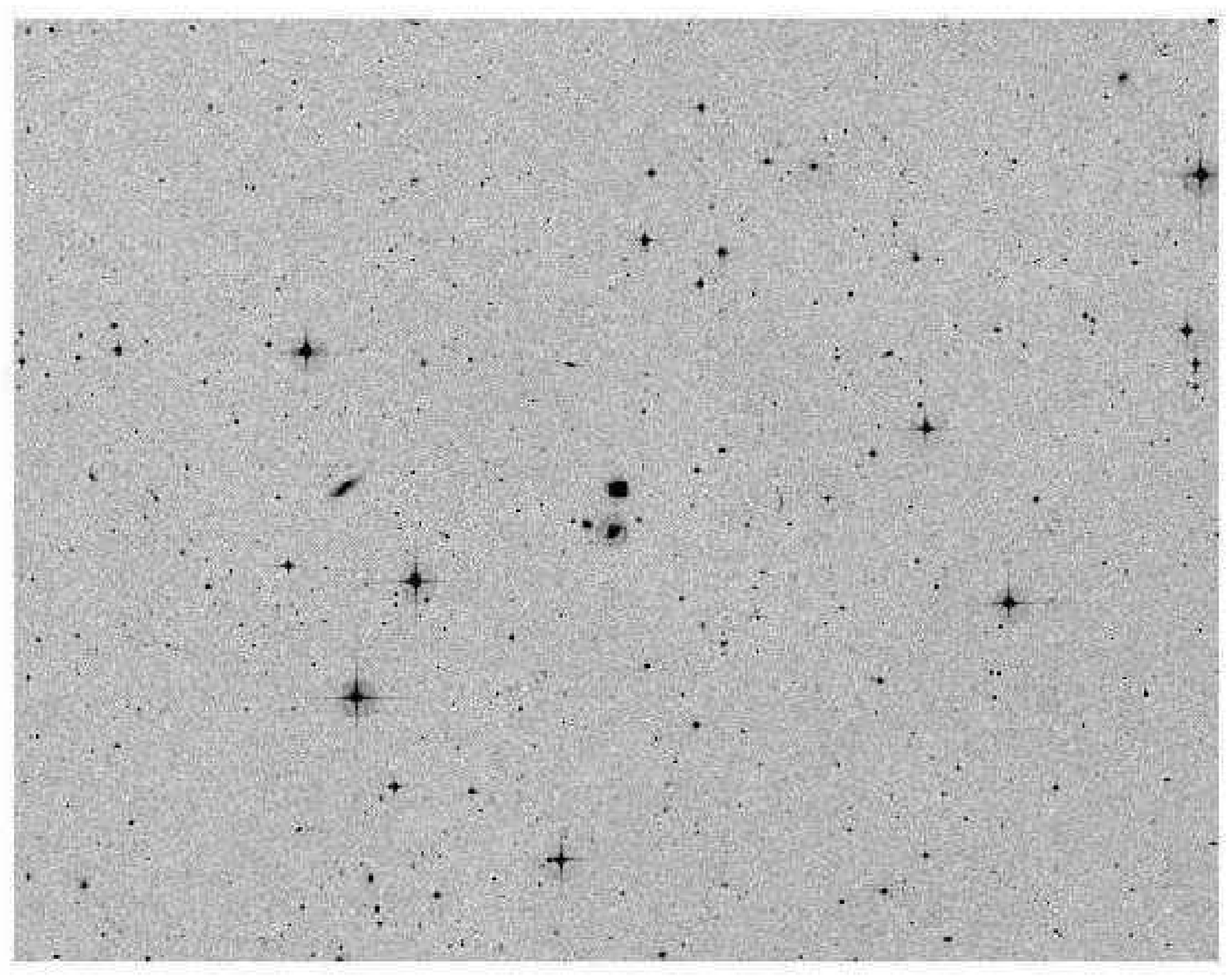} \hskip 1cm 
  \epsfxsize=8.0cm \epsfbox{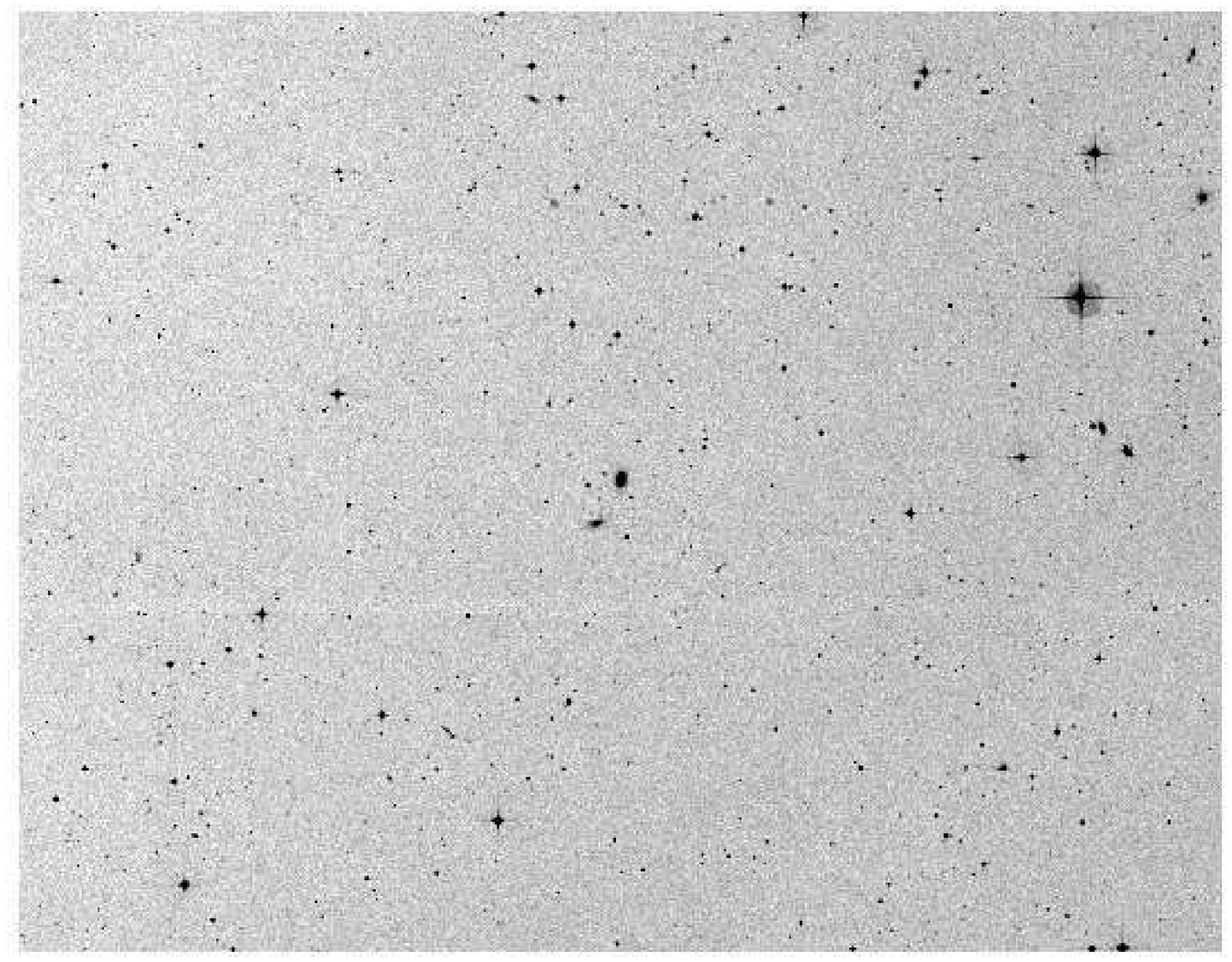}} \\
\centerline{VLG031 \hskip 6.6cm VLG040}
  {\epsfxsize=8.0cm \epsfbox{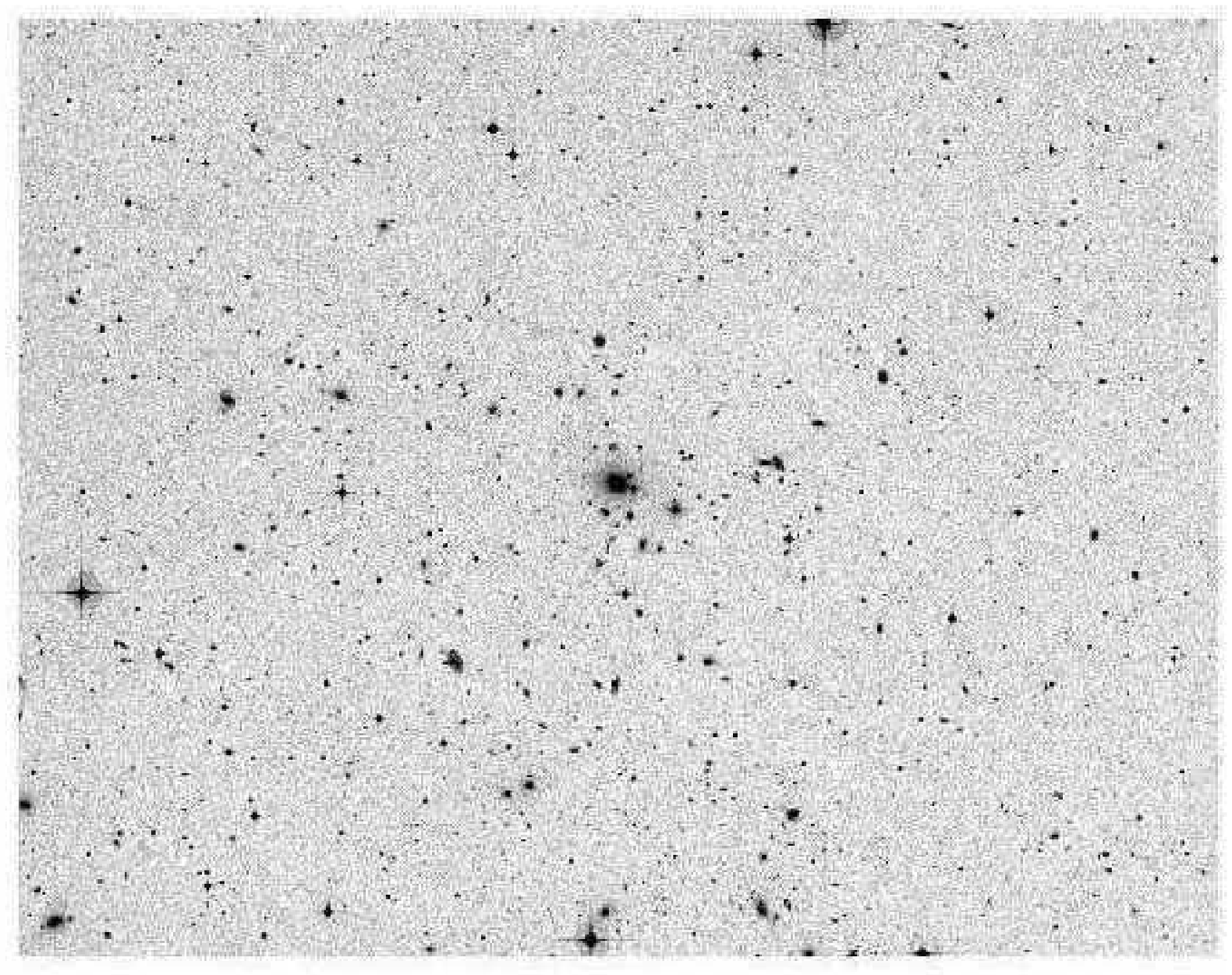} \hskip 1cm 
  \epsfxsize=8.0cm \epsfbox{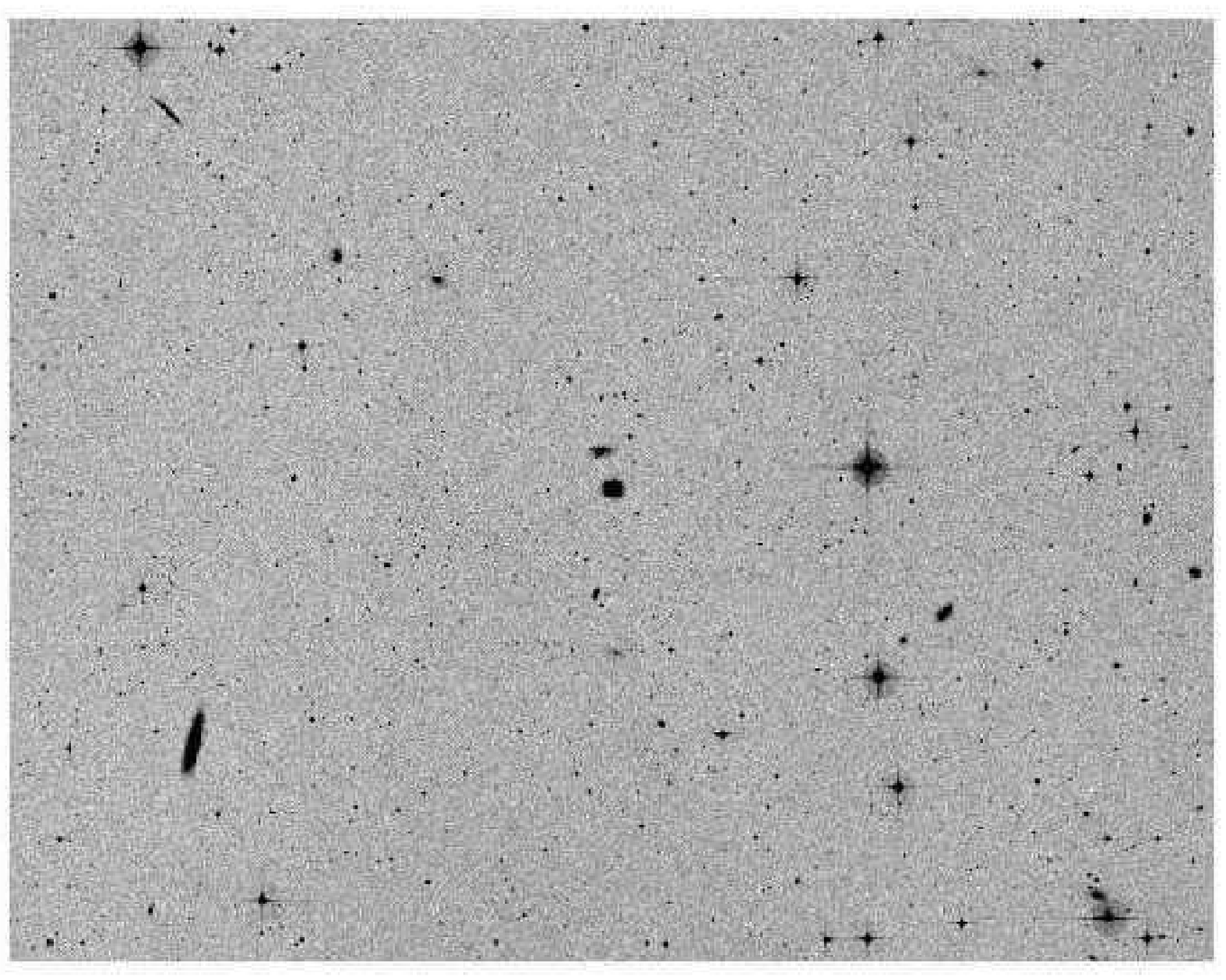}} \\
%\end{figure}
%\begin{figure*}
%\caption[]{\em{(cont.)} Finding charts 
% of the VLG fields selected from the 2dFGRS.}
 \centerline{VLG043 \hskip 6.6cm VLG045}
 {\epsfxsize=8.0cm \epsfbox{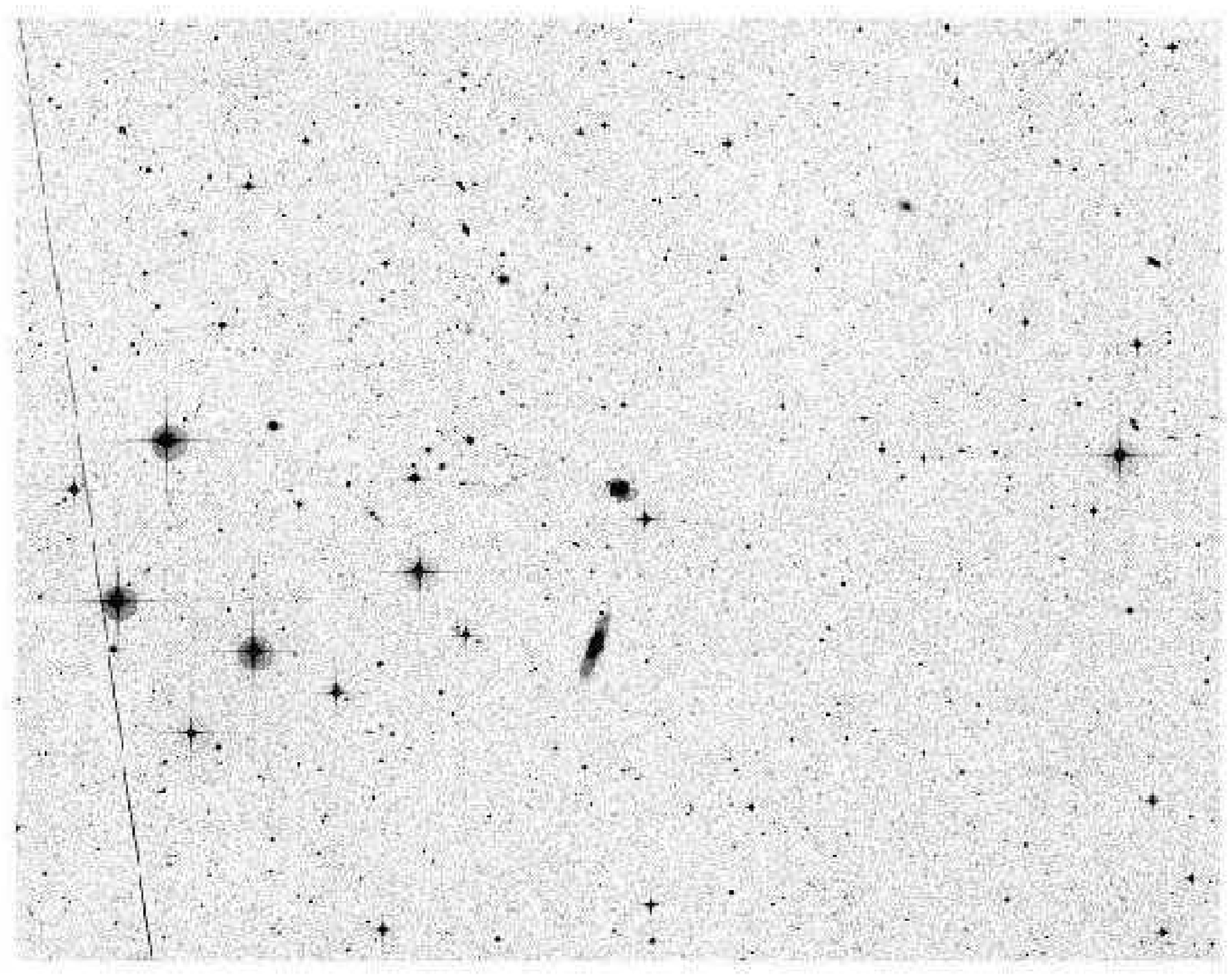} \hskip 1cm 
  \epsfxsize=8.0cm \epsfbox{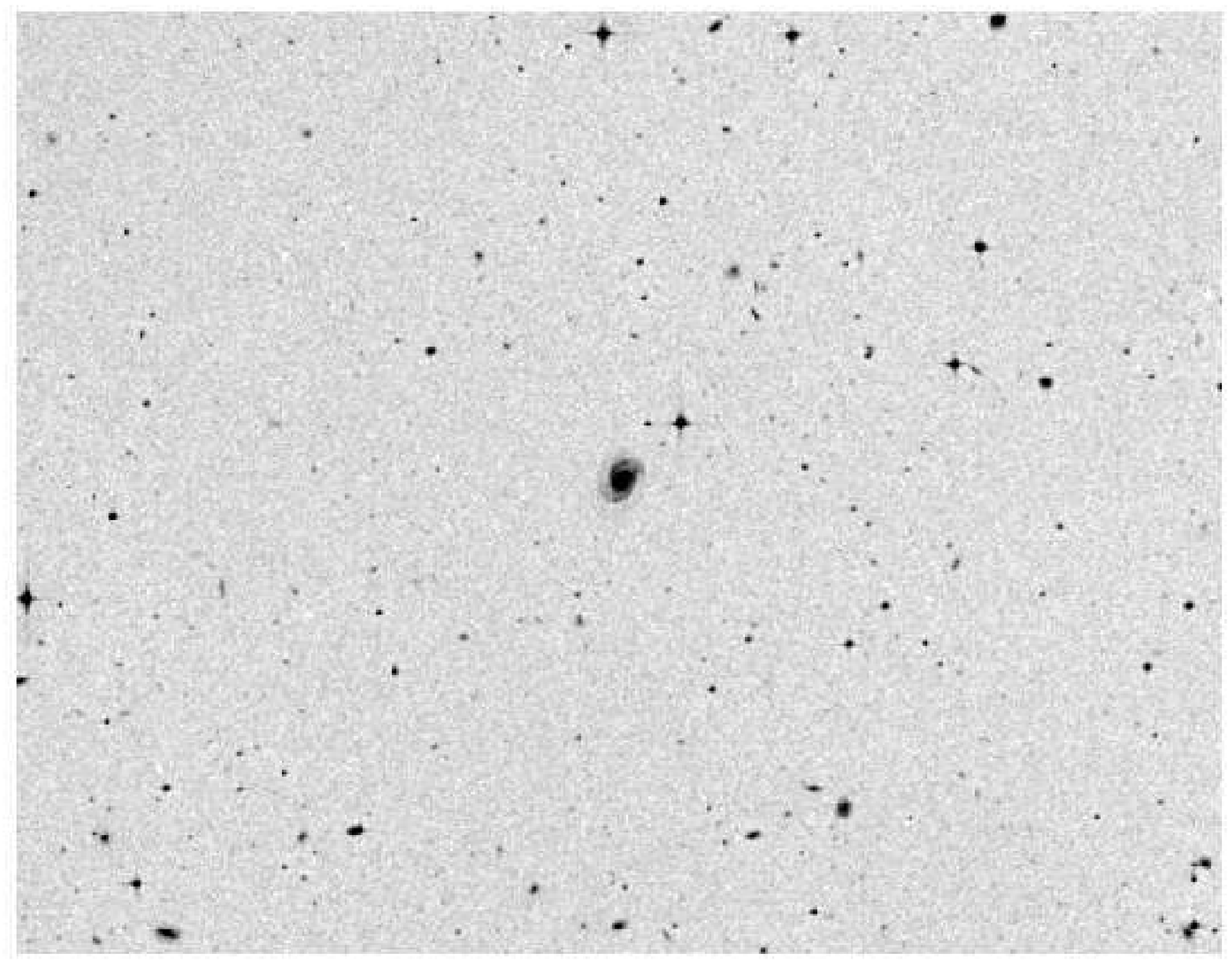}} \\
\end{figure}
\begin{figure*}
%\vskip 1.65cm
%\caption[]{Fields of the VLG fields selected from the 2dFGRS.}
\centerline{VLG048 \hskip 6.6cm VLG053}
 {\epsfxsize=8.0cm \epsfbox{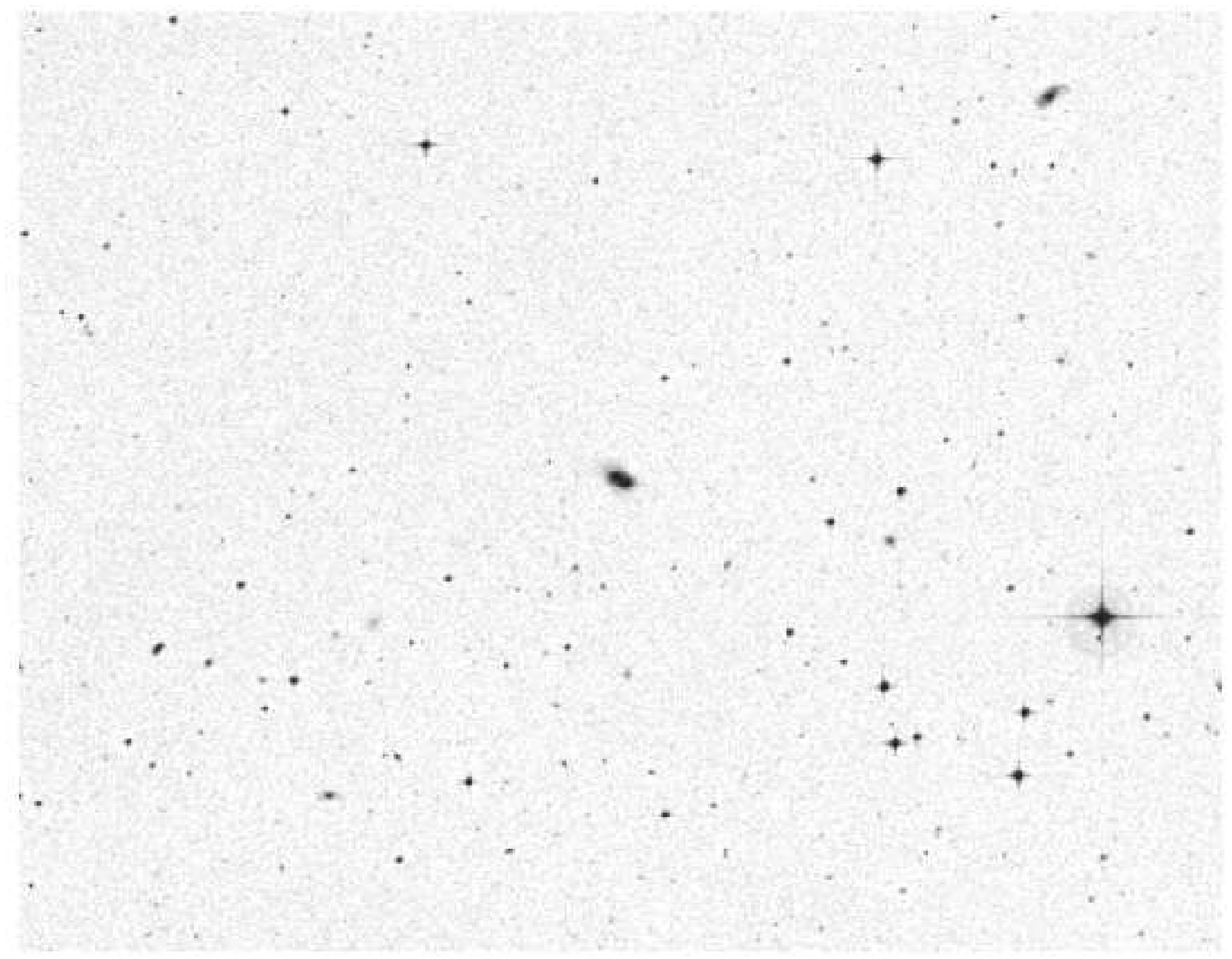} \hskip 1cm 
  \epsfxsize=8.0cm \epsfbox{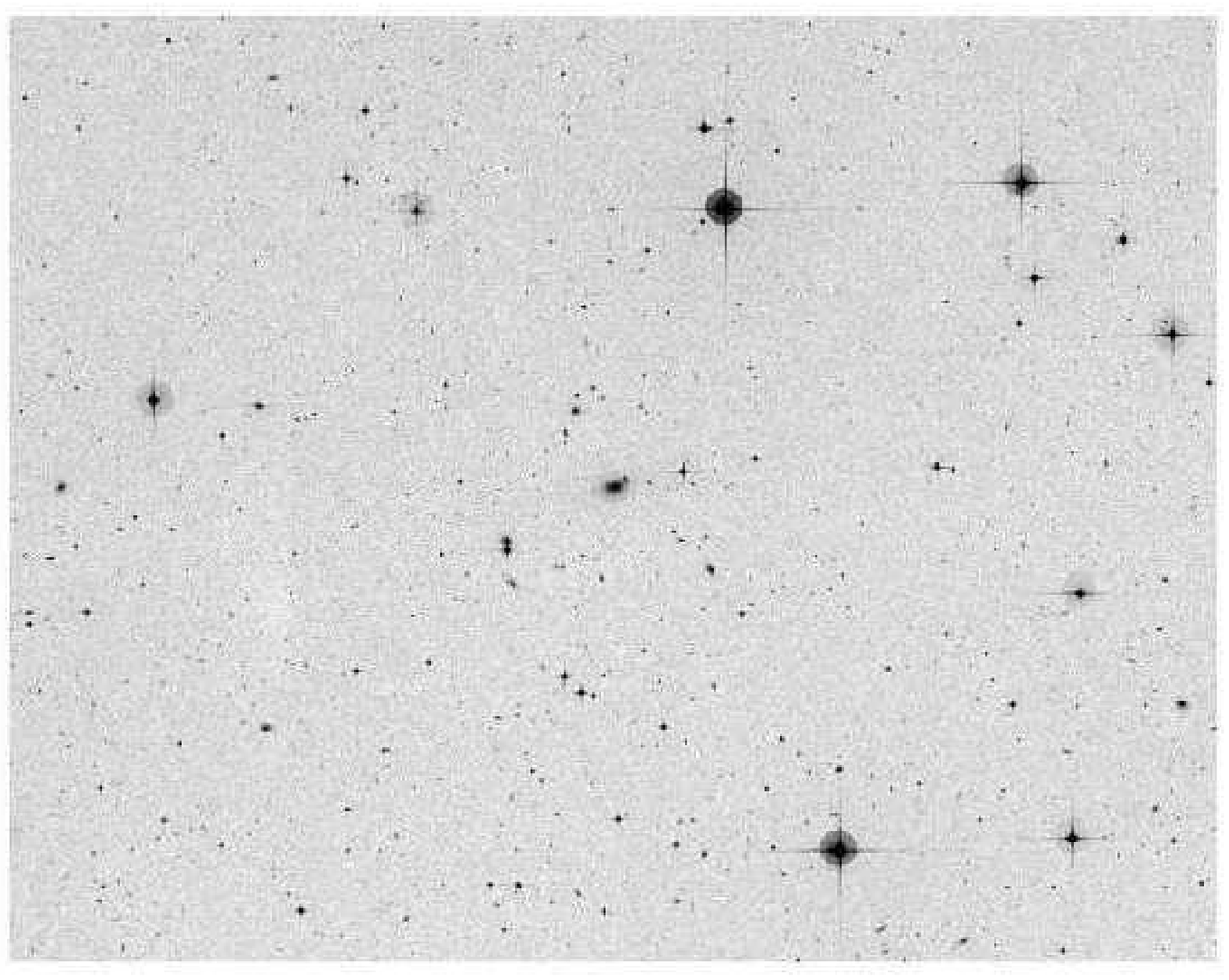}} \\
\end{figure*}
\begin{figure*}
\centerline{VLG069 \hskip 6.6cm VLG083}
 {\epsfxsize=8.0cm \epsfbox{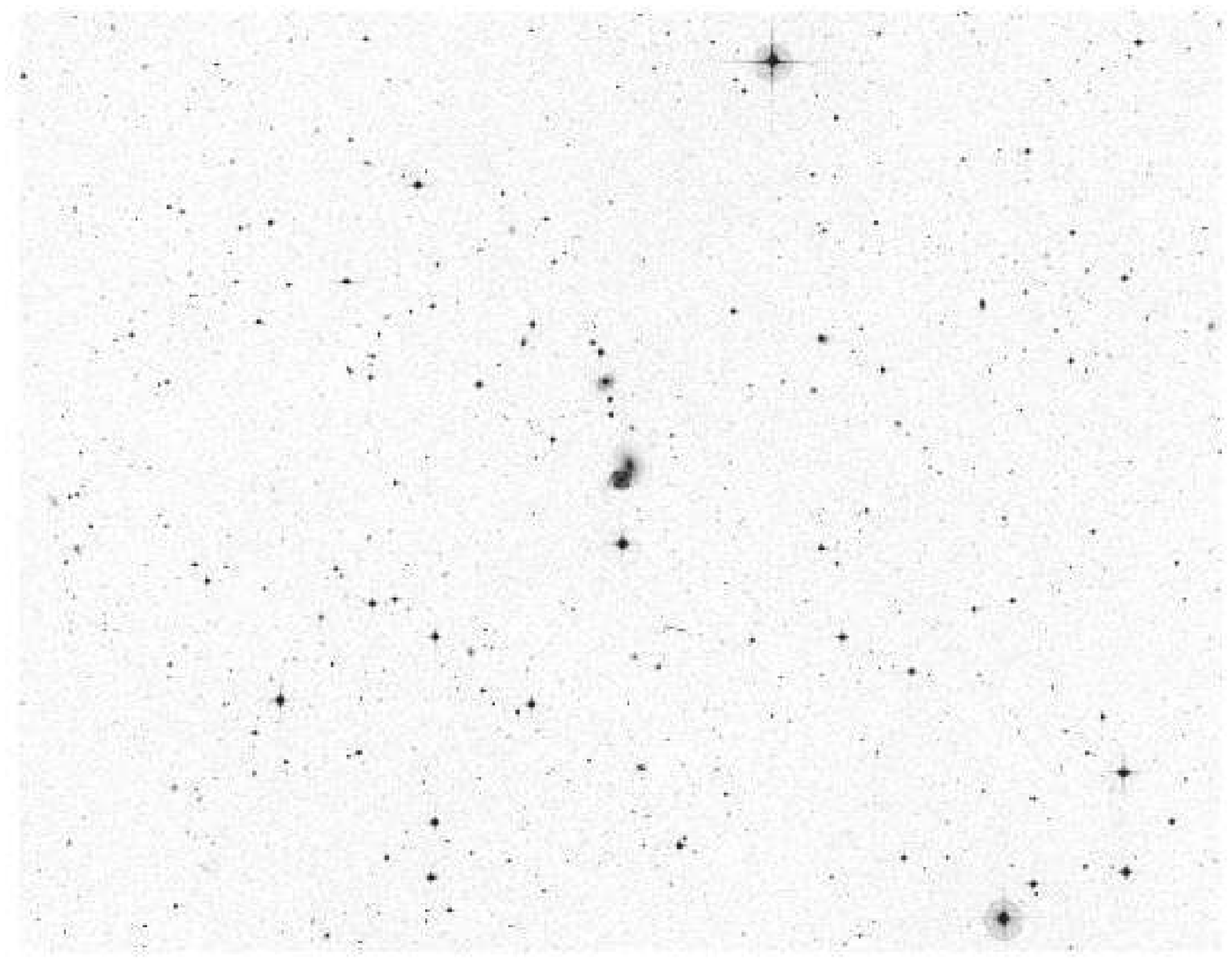} \hskip 1cm 
  \epsfxsize=8.0cm \epsfbox{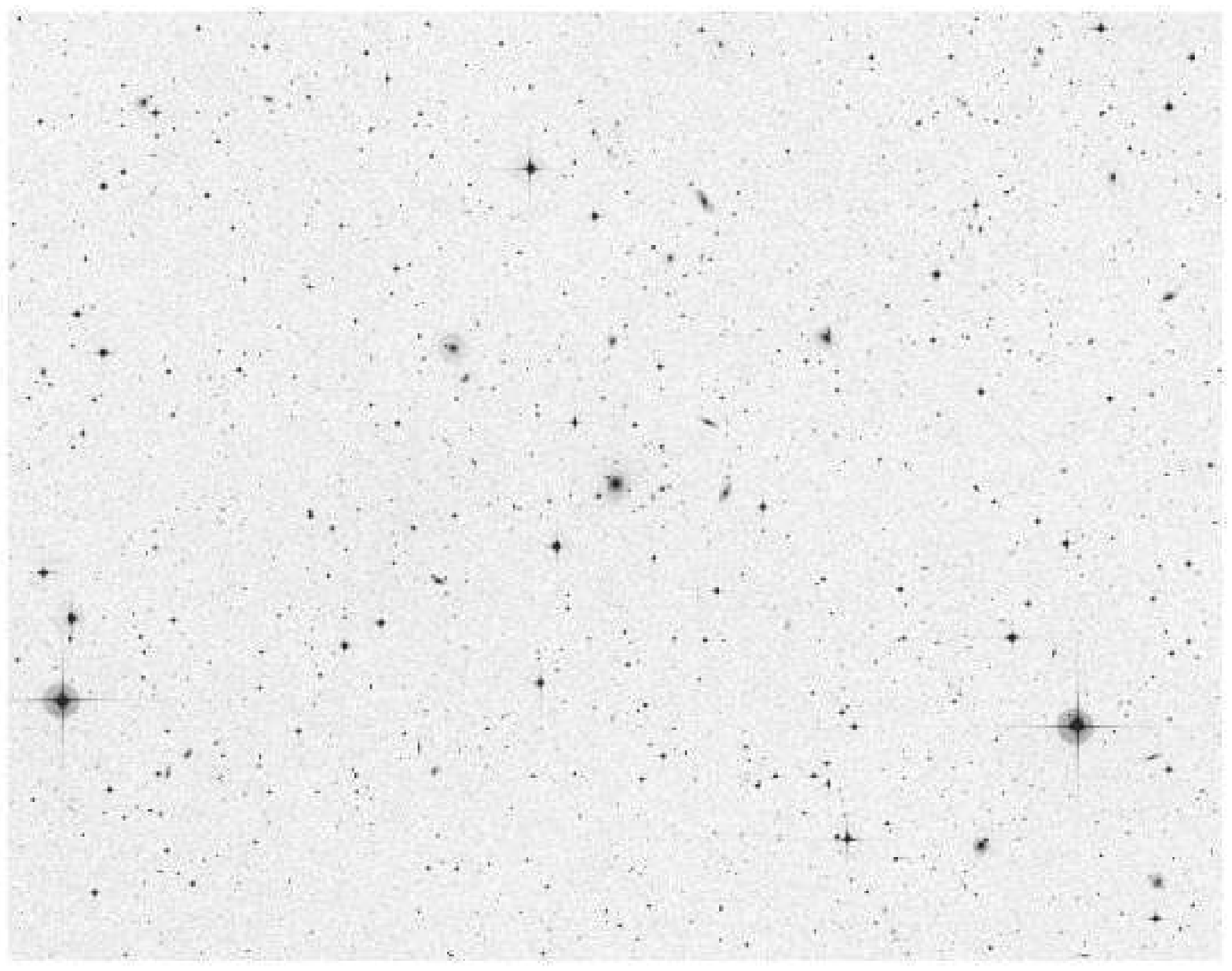}} \\
\centerline{VLG086 \hskip 6.6cm VLG093}
 {\epsfxsize=8.0cm \epsfbox{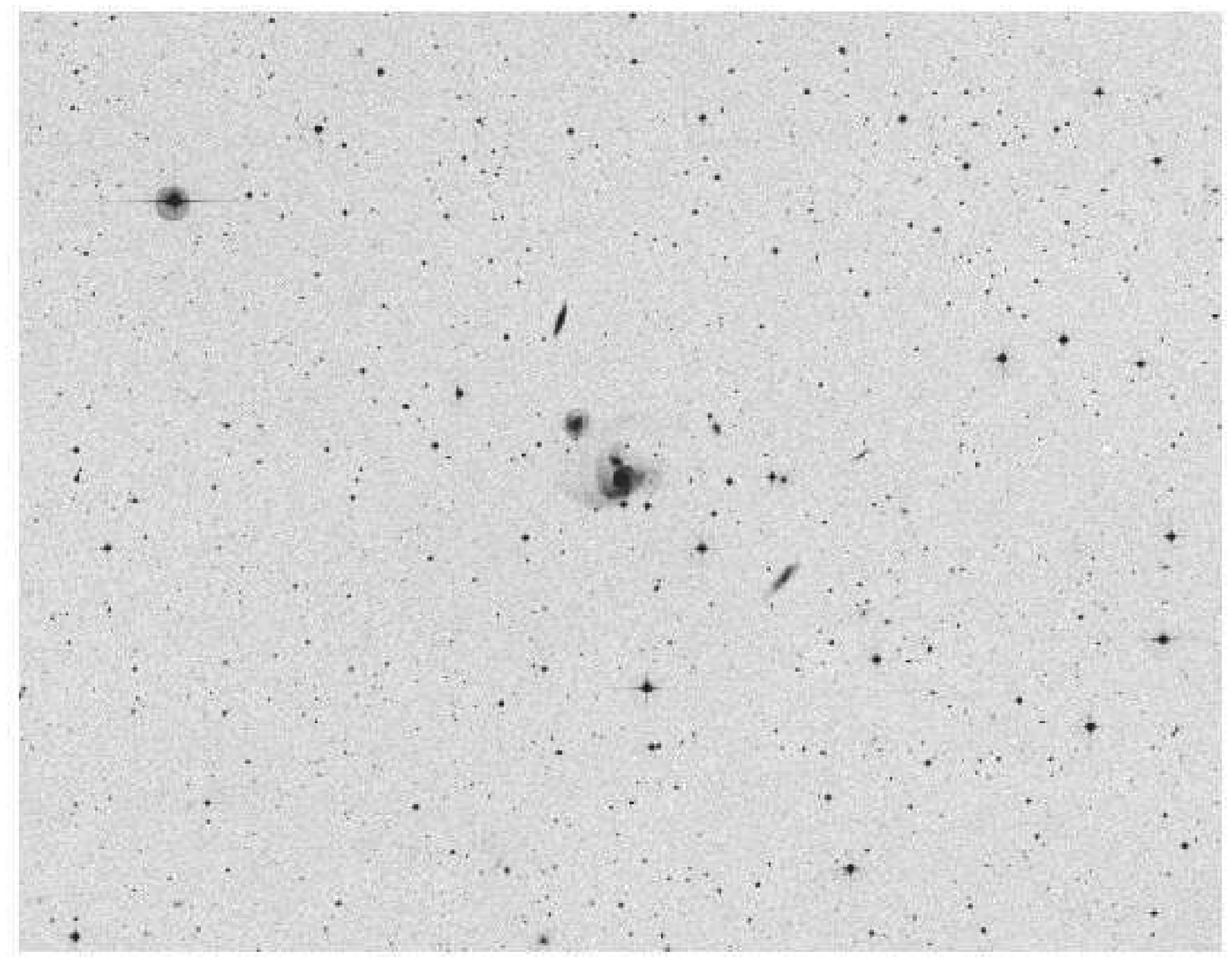} \hskip 1cm 
  \epsfxsize=8.0cm \epsfbox{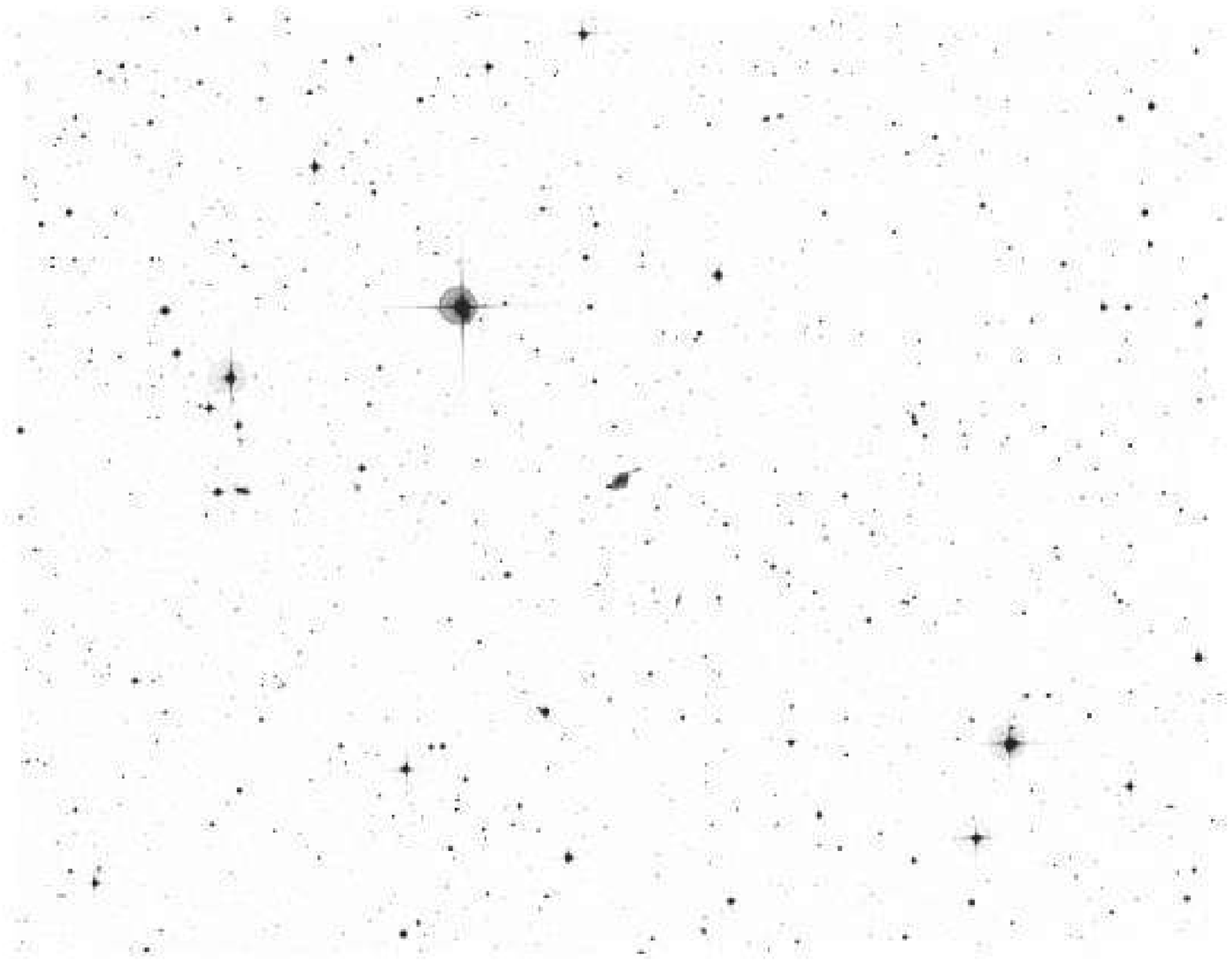}} \\
\end{figure*}
\begin{figure*}
%\caption[]{Finding charts of the VLG fields selected from the 2dFGRS.}
\centerline{VLG094 \hskip 6.6cm VLG108}
 {\epsfxsize=8.0cm \epsfbox{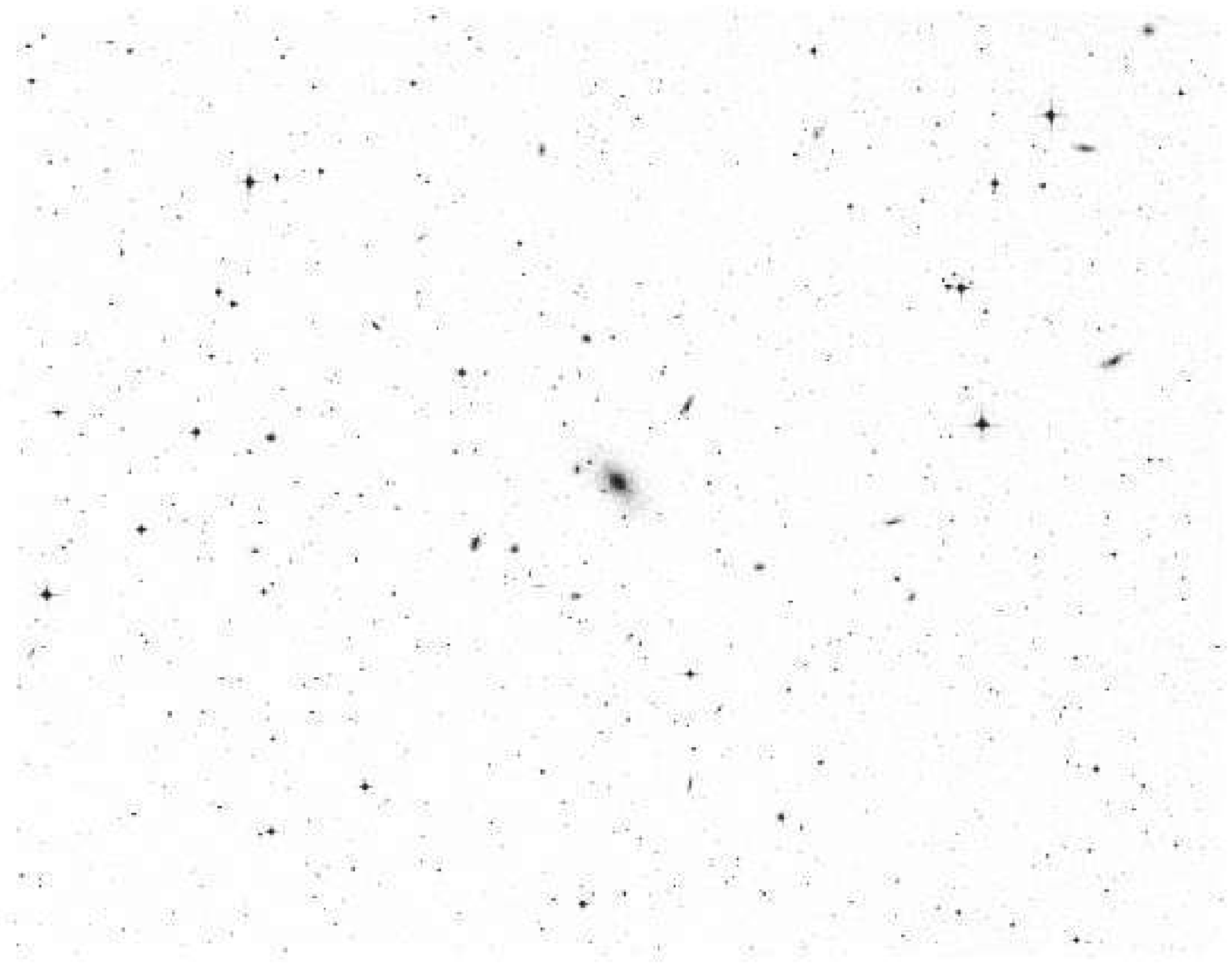} \hskip 1cm 
  \epsfxsize=8.0cm \epsfbox{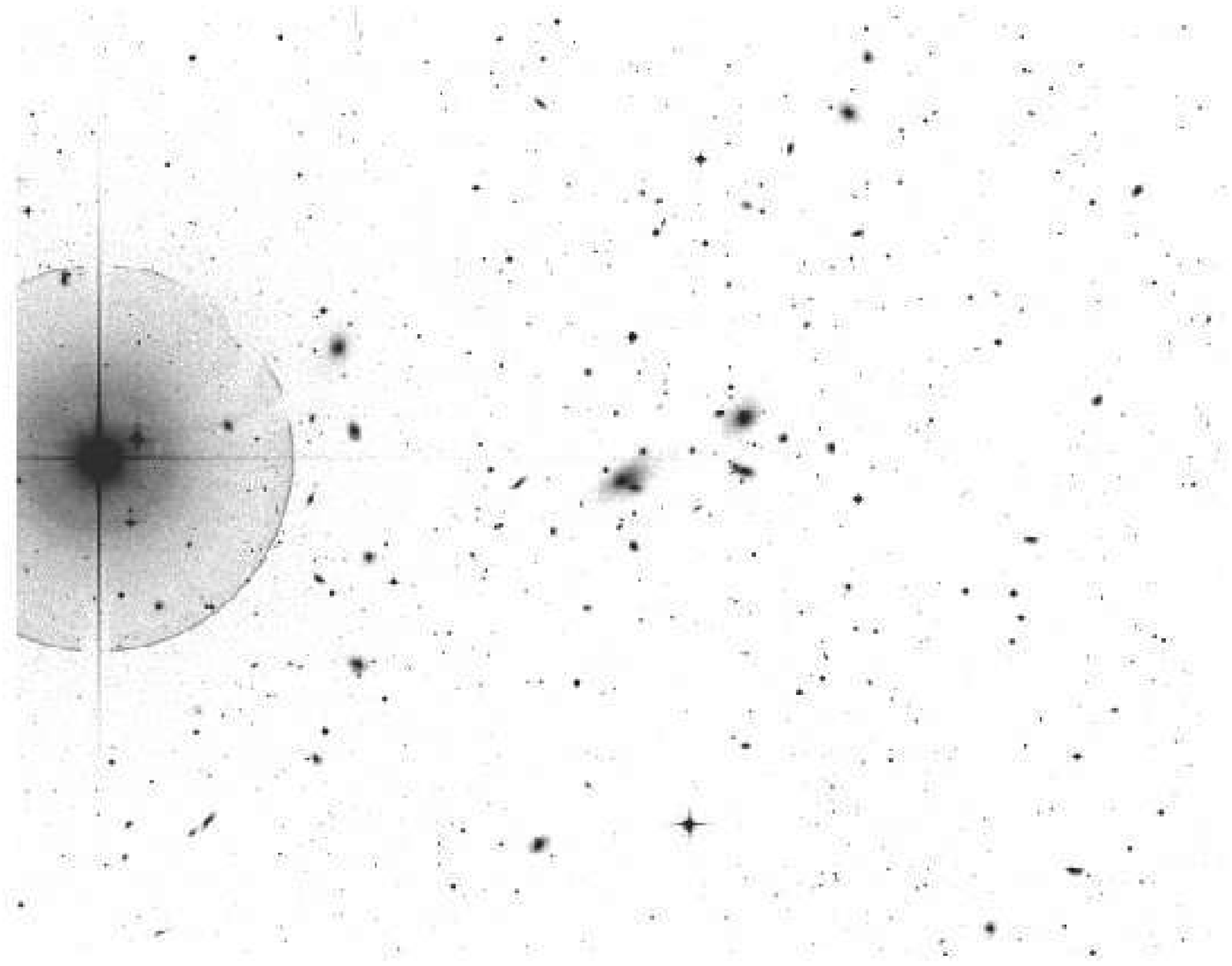}} \\
\centerline{VLG109 \hskip 6.6cm}
 {\epsfxsize=8.0cm \epsfbox{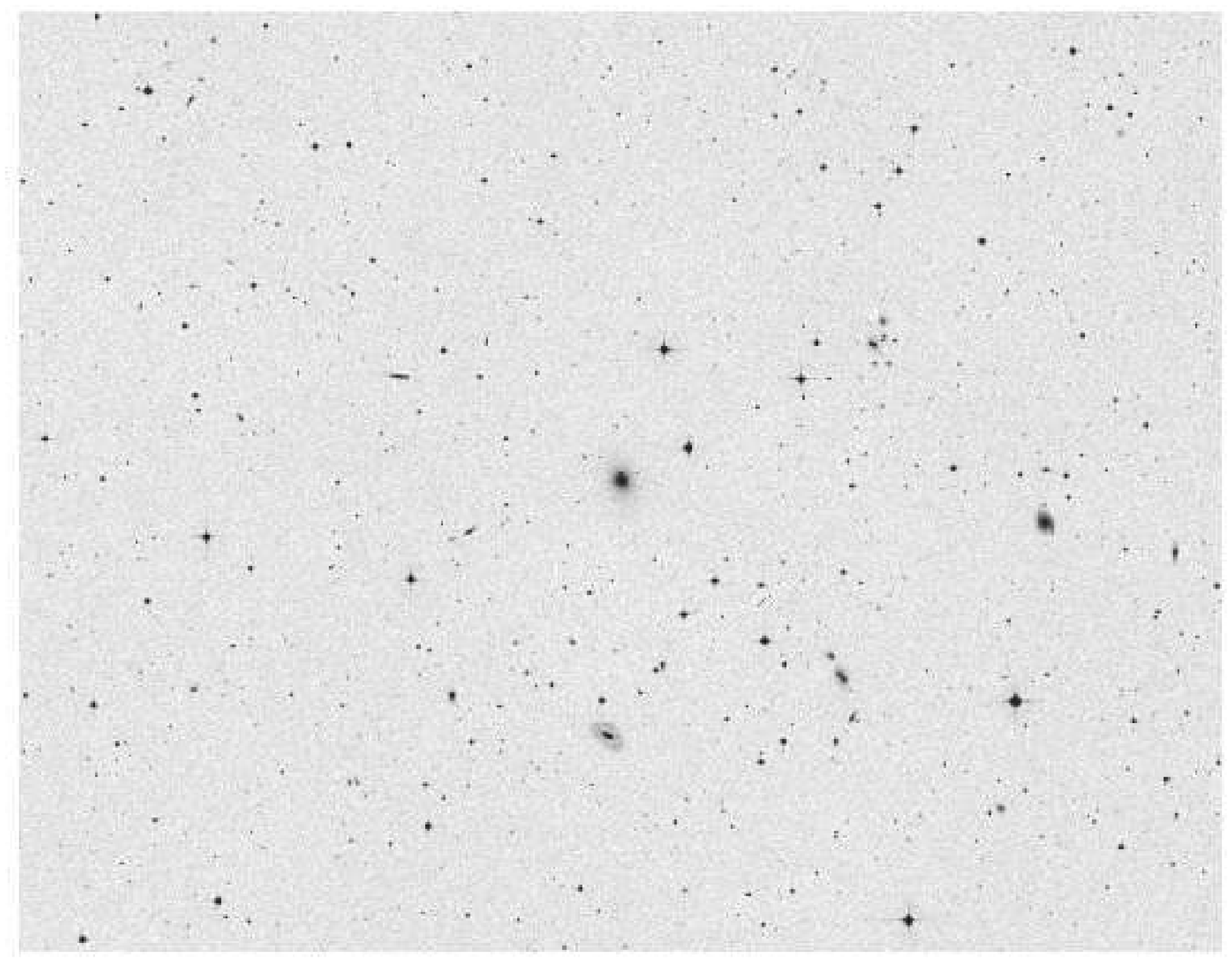} \hskip 1cm} \\
\end{figure*}

\subsection{Systems observed at OHP}
 
At the Observatoire de Haute--Provence we could observe only galaxies
at $\delta > -10^o$, while SSRS2 galaxies are south of 
$\delta = -2.5^o$. Among the 12 VLGs in our catalogue which satisfied 
this declination constraint, we chose three VLGs around which 
an inspection of DSS images had revealed the presence
of galaxies fainter 
than the VLG but still bright enough to get a useful spectrum at
the 1.93m telescope. 
The selected VLGs are VLG061, VLG068 and VLG074.
In order to cover the whole night, we included also 
a VLG galaxy selected from the CfA catalogue (VLG 0716+5323), and Arp 127,
a galaxy we had included in a preliminary version of the VLG catalogue but
finally excluded for its discrepant redshift and the consequent uncertainty
on its absolute magnitude. This system is presented in appendix A.

Our observations were carried out in 1997 at the 1.93m telescope 
with the Carelec spectrograph in long slit mode at the Cassegrain focus. 
The grating dispersion was 260\AA/mm, corresponding to $\sim 7$~ \AA
with the 512$\times$512 pixels of the Tektronix CCD.
Data reduction was performed with IRAF; calibrations were done
using the OHP He--Ar lamps.
Redshifts were measured with
{\em xcsao} in the {\em rvsao} package, using 5 star templates and 
attributing to each galaxy the redshift given by the best--fitting template 
(i.e. the one with the highest $R$ parameter, see Tonry \& Davis 1979).
One velocity standard star and one galaxy with velocity 
measured from HI observations were also observed and used as a check of the 
zero--point calibration, cross--correlating them with a subset of our spectra
and with our templates: 
in both cases, the redshifts were consistent within 10 km/s.

Positions and redshifts of the observed galaxies are listed in table 1.

\begin{table*}[ht]
 \label{tab:observations}
 \caption[]{Heliocentric Redshifts of Galaxies in VLG fields}
 \begin{flushleft}
 \begin{tabular}{rrrrrl}
  \hline
   \hline
 Iden.  &   RA (J2000) & DEC (J2000) &  $V_h$ (km/s) & Error & Notes \\
 \hline
%--------------------------------------------------------------- start data
 VLG061      &  02 30 42.7 & -02 56 21 & 5719 & 31 &                \\
 1           &  02 30 32.1 & -02 53 07  &  5950   &  24 & Emission lines \\ % B
 2           &  02 30 44.7 & -02 53 58  &  5488   &  36 &                \\ % A
 3           &  02 30 46.4 & -02 57 06  &  5561   &  28 &                \\ % C
 4           &  02 30 47.6 & -02 54 32  &  5870   &  20 & Emission lines \\ % E
 5           &  02 30 48.9 & -02 56 46  &  5189   &  91 &                \\ % D
 6           &  02 31 11.6 & -02 56 35  &  5847   &  31 &  m=15.76   \\ % F
 7           &  02 30 21.9 & -02 59 07  & 12498   &  56 &             \\ % G
\hline
 VLG068 &  03 25 11.5 & -06 10 52  &  9933   &      48 & 
V=10107   m=13.97           \\
 MRK0609  &  03 25 25.3 & -06 08 38  & 10264   &      57 & V=10236       \\
 MRK0610  &  03 25 31.4 & -06 07 43  & 10408   &      55 & V=10301       \\
\hline
 VLG074a   & 04 08 07.4 & -08 49 45  & 9930 & 50 & NGC1516A \\
 VLG074b   & 04 08 08.2 & -08 50 06  & 9864 & 45 & NGC1516B \\
 1         & 04 07 45.3 & -08 44 26  & 10073   &      37 &               \\ % B
 2         & 04 07 59.0 & -08 50 24  & 36482   &      61 &           \\ % A
 3         & 04 08 00.3 & -08 49 24  & 46248   &      65 & $H_\beta$, 
[OIII]4959 \& 5007 \\ % D
 4         & 04 08 06.4 & -08 48 04  & 36342   &     101 &           \\ % C
 5         & 04 08 12.1 & -08 56 11  & 42342   &      95 &           \\ % G
 6         & 04 08 21.1 & -08 47 20  & 36254   &      45 &           \\ % E
 \hline
%--------------------------------------------------------------- start data
VLG 0716+5323  & 07 16 41.2 & 53 23 09 & 19069 & 41 & m=14.0, V=19307, 
X--ray  \\
 1               & 07 16 19.9 & 53 21 51 &   137 & 17 & Star \\
 2               & 07 16 21.2 & 53 21 59 & 19048 & 32 & \\
 3               & 07 16 24.4 & 53 21 37 & 19573 & 25 & \\
 4               & 07 16 32.0 & 53 23 45 & 18330 & 44 & \\
 5               & 07 16 38.1 & 53 15 38 & 19781 & 47 & \\
 6               & 07 16 40.0 & 53 22 23 & 19904 & 43 & \\
 7               & 07 16 43.1 & 53 22 55 & 20522 & 73 & \\
 8               & 07 16 47.9 & 53 22 45 & 19066 & 57 & \\
 9               & 07 17 29.2 & 53 24 45 & 19043 & 42 & \\ 
%--------------------------------------------- end data
\hline
\end{tabular}
\end{flushleft}
\end{table*}

\subsection{2dFGRS data}

Part of the SSRS2 region is covered by the 2dFRS
(see Colless et al. 2001), and we searched for
galaxies around VLG positions 
in the presently available public catalogue (the ``100k'' 
catalogue\footnote{electronically available at the address 
 {\em http://msowww.anu.edu.au/2dFGRS/Public/main.html}},
including more than 102,000 redshifts).

The limiting magnitude of the 2dFGRS is $b_J=19.45$; 
within the maximum distance defined by the SSRS2 VLG volume--limited sample
($\sim 0.065$), the 2dFGRS is volume--limited at $M=-17$ 
(of course the field incompleteness has to be taken into account).

We selected all galaxies in the 2dFGRS within a projected separation less than 
1.5\h~ and a velocity difference less than 1500 km/s with respect to the 
SSRS2 VLGs. The velocity cut was chosen to limit foreground and background 
contamination, but large enough to include also marginal members. 
We also applied a 3--$\sigma$ clipping (see e.g. Yahil \& Vidal 1977) to the 
velocity distribution. 
The chosen value of the projected radius corresponds to one Abell radius and
it is also used as a criterion for determining the Local Group membership
(van den Bergh 1999). We found data for 19 VLG fields: in 4 cases
(VLG003, VLG008, VLG054, VLG075) we could retrieve only one redshift 
(in addition to the VLG) with our selection limits.
Such cases deserve a more careful study and we will not include them in the 
present work. We simply note here that VLG075 is in a group 
identified in the Las Campanas Redshift Survey (Tucker et al. 2000).
In the other 15 fields we could obtain at least 5 redshifts
(reduced to 3 by the 3$\sigma$ clipping in the case of VLG022). 

The images of the fields centered on the selected VLGs
(30 arcmin size in $\delta$) were retrieved from the 
Digitized Sky Survey and are shown in figures \ref{fig:ohpfields}
and \ref{fig:fields}.
In appendix B and in tables available in electronic
form we give positions and redshifts of the 2dFGRS galaxies selected
according to the criteria defined above.

\section{Properties of the VLG systems}

We have calculated the mean redshift and the velocity dispersion for each
VLG system; the main properties are shown in table
\ref{tab:systems}, where we
list in column (1) the VLG number, in column (2) the VLG morphological type,
in column (3) the total number of galaxies after applying
the 3--$\sigma$ clipping and used for measuring the mean redshift and velocity
dispersion, in column (4) the mean heliocentric redshift
with its error, in column (5) the velocity dispersion,
in column (6) the system type, when available from the literature.

With respect to the known systems listed in table 2 of our paper I, 
we could associate other 8 VLGs to galaxy systems, and
increase the number of measured velocity
dispersions: in fact only VLG086 (in an Hickson compact group) and VLG108 
(in the ACO cluster A4038) had already an estimate of the velocity
dispersion.

\begin{table*}[ht]
\caption[]{Properties of the VLG systems}
  \label{tab:systems}
 \begin{flushleft}
 \begin{tabular}{rlrcrl}
  \hline
   \hline
 Ident. & Type & $N_z$ & $\bar{z}$ & $\sigma_r$ (km/s) & System Type \\
 \hline
%--------------------------------------------------------------- start data
 14 & Sbc      & 14 & $0.0389 \pm 0.0007$ & $747 _{~-114} ^{~+204}$ & 
Triplet   \\
 22 & S0       &  3 & $0.0570 \pm 0.0003$ & $137 \pm 66$            & SCG55  \\
 31 & D        & 15 & $0.0536 \pm 0.0007$ & $720 _{~-107} ^{~+188}$ & A151   \\
 40 & S        & 10 & $0.0403 \pm 0.0008$ & $724 _{~-127} ^{~+256}$ & ---    \\
 43 & Sb       & 15 & $0.0567 \pm 0.0004$ & $468 _{~-72}  ^{~+123}$ & ---    \\
 45 & S        & 21 & $0.0572 \pm 0.0004$ & $464 _{~-62}  ^{~+97}$  & ---    \\
 48 & Sbc      & 20 & $0.0584 \pm 0.0003$ & $378 _{~-53}  ^{~+82}$  & Binary \\
 53 & S0       & 33 & $0.0575 \pm 0.0002$ & $360 _{~-41}  ^{~+57}$  & ---    \\
 61 & SB(rs)c  &  7 & $0.0193 \pm 0.0004$ & $318 _{~-67}  ^{+177}$  & ---    \\
 68 & S Sy1    &  3 & $0.0340 \pm 0.0004$ & $236 \pm 99$       & SSRS2 group \\
 69 & SB(s)b p &  8 & $0.0376 \pm 0.0002$ & $158 _{~-46}  ^{+79}$ & Binary \\ 
 74 & S        &  3 & $0.0332 \pm 0.0003$ & $104 \pm 56$          & Binary \\
 83 & E        & 31 & $0.0580 \pm 0.0002$ & $319 _{~-38} ^{~+53}$ & S0983    \\
 86 & SB(s)bc p:Sy & 37 & $0.0231 \pm 0.0004$ & $669 _{~-68} ^{~+96}$ & HCG91\\
 93 & Sb       & 15 & $0.0566 \pm 0.0005$ &  $550 _{~-83} ^{~+144}$ & ---    \\
 94 & SAB(rs)p & 39 & $0.0336 \pm 0.0002$ & $413 _{~-43} ^{~+59}$ & EDCC155  \\
108 & cD       & 114 & $0.0297 \pm 0.0002$ & $659 _{~-41} ^{~+49}$ & A4038   \\
109 & E        & 13 & $0.0492 \pm 0.0003$ & $297 _{~-52} ^{~+89}$ & S1155 \\ 
VLGN 0716+5323 & E &~9 & $0.0645 \pm 0.0007$ & $609 ^{+234} _{-109}$ & Z1261 \\
%--------------------------------------------------------------- end data
\hline
\end{tabular}
\end{flushleft}
\end{table*}

\begin{figure*}[ht]
\caption[]{Redshift histograms of the VLG systems with $\ge 10$ redshifts 
(dashed line: VLG redshift).}
\label{fig:hist}
{\epsfxsize=4.1cm \epsfbox{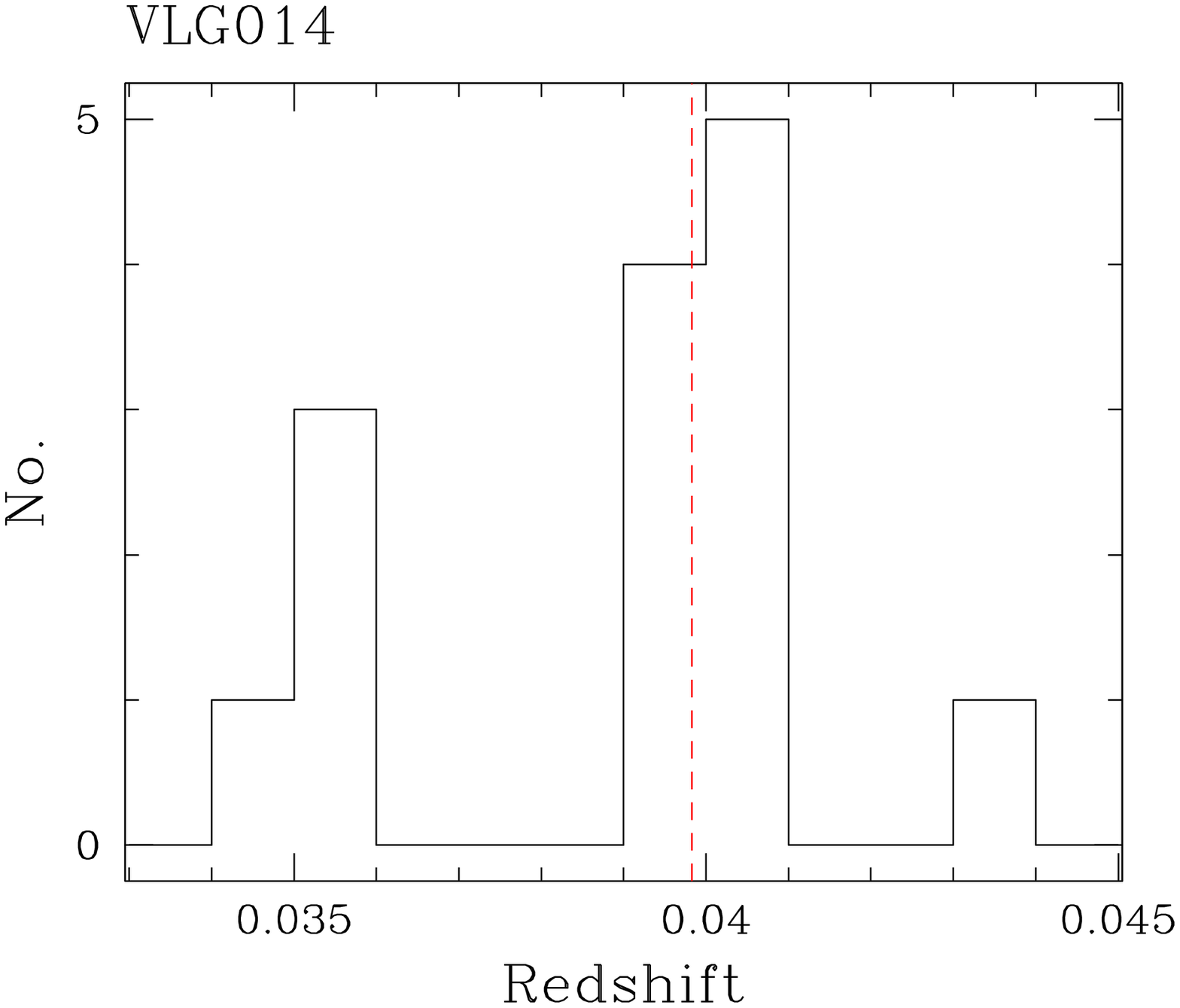} \hskip 0.1cm 
 \epsfxsize=4.1cm \epsfbox{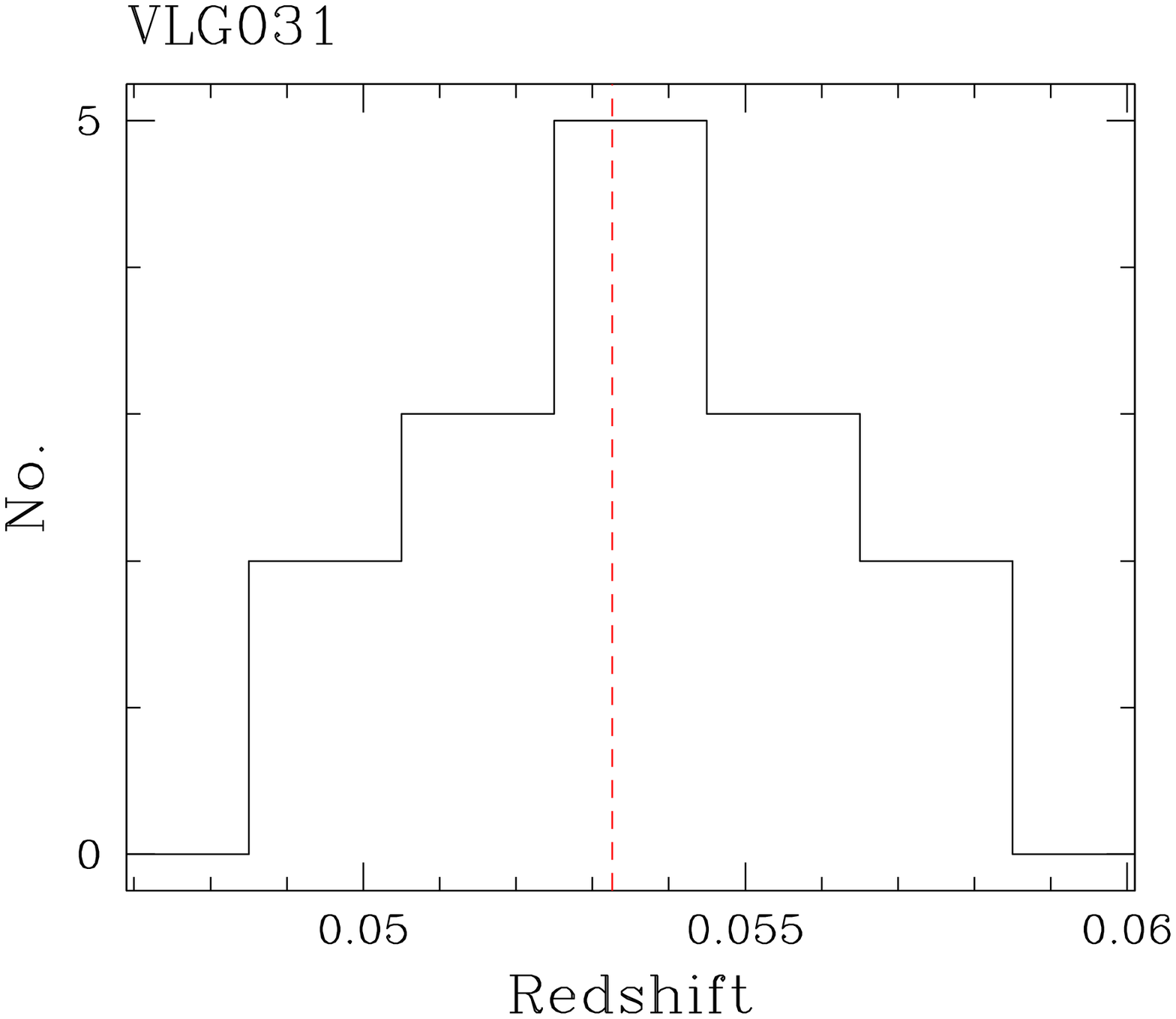} \hskip 0.1cm
 \epsfxsize=4.1cm \epsfbox{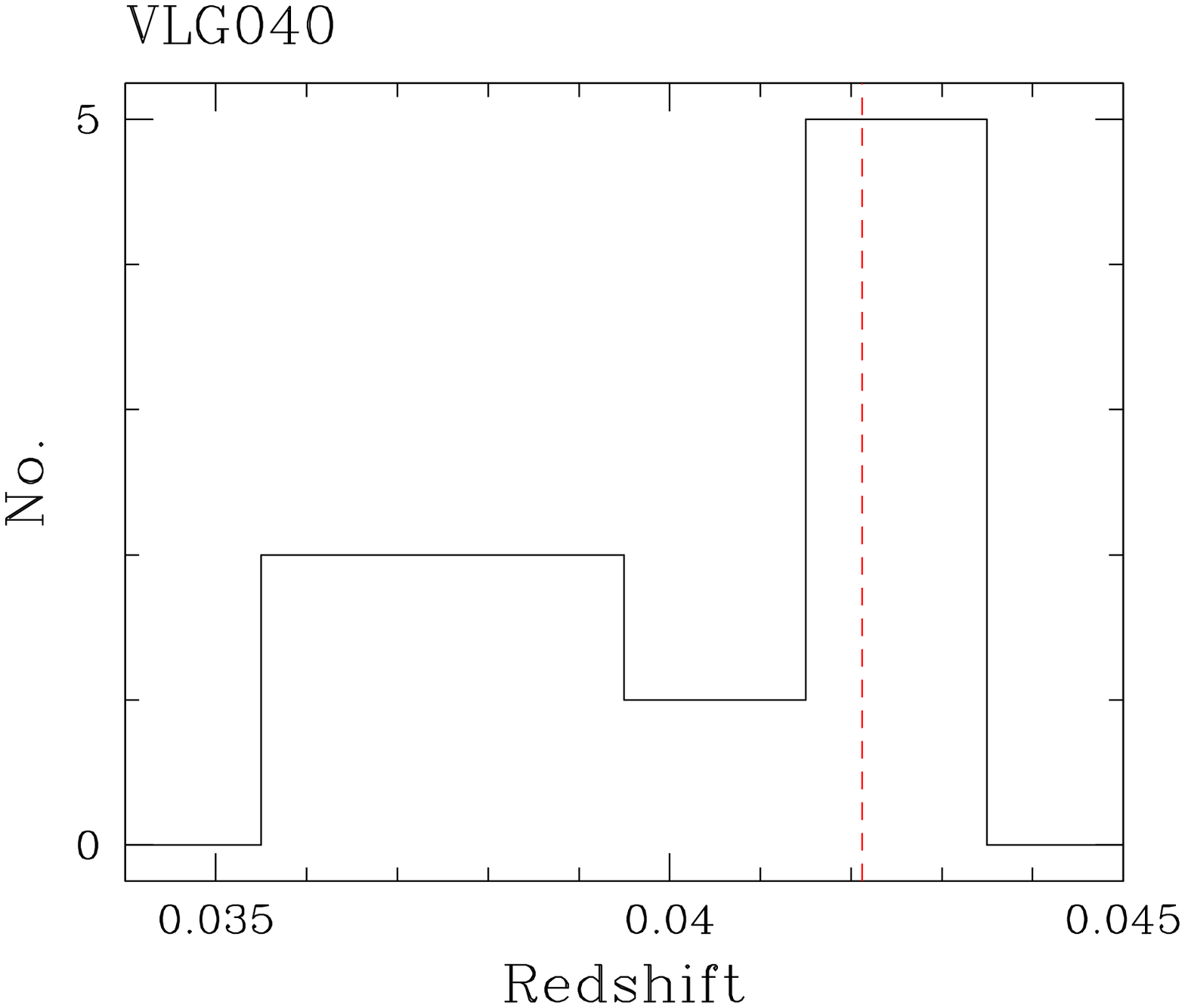}} \\ 
{\epsfxsize=4.1cm \epsfbox{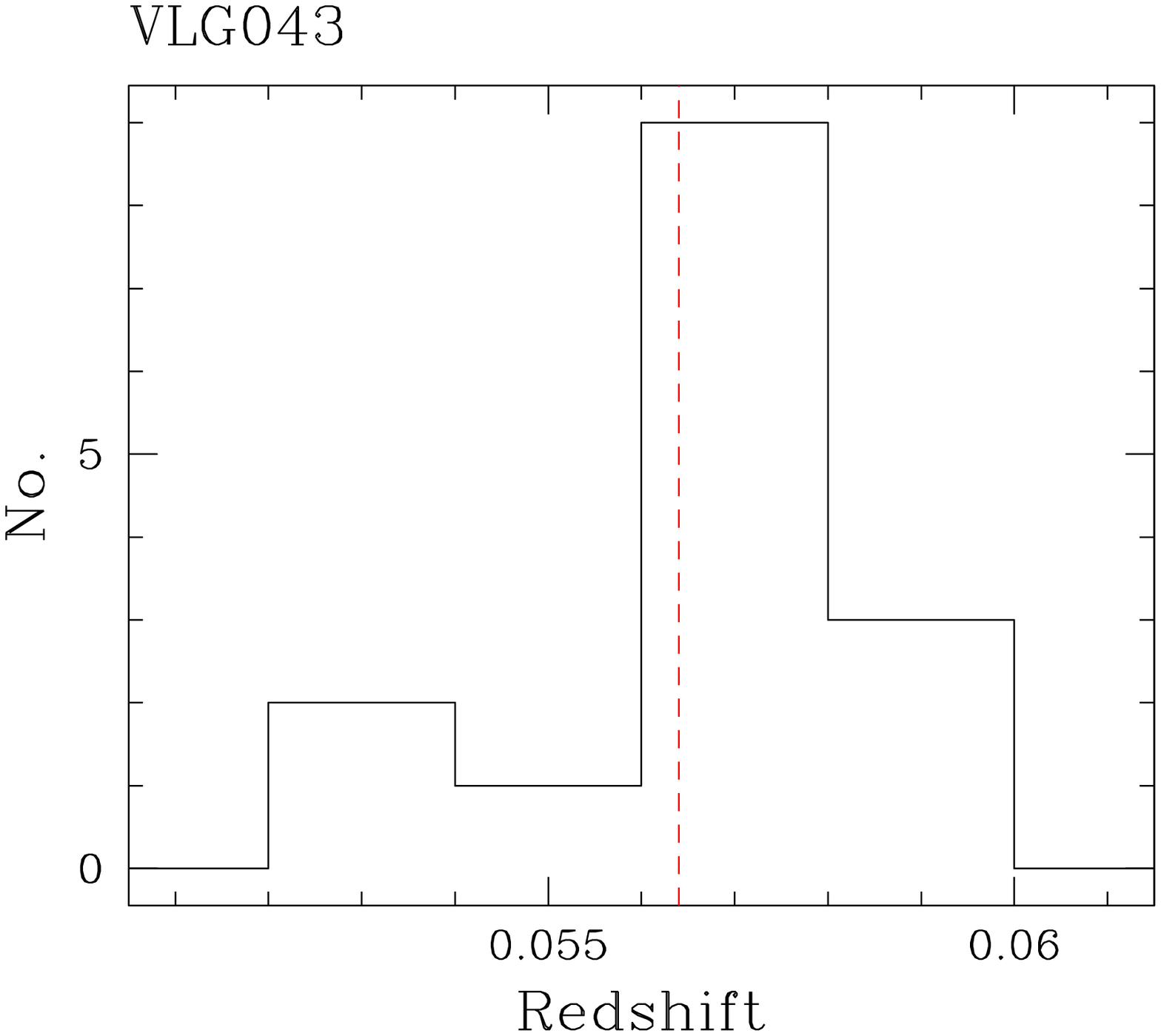} \hskip 0.1cm 
 \epsfxsize=4.1cm \epsfbox{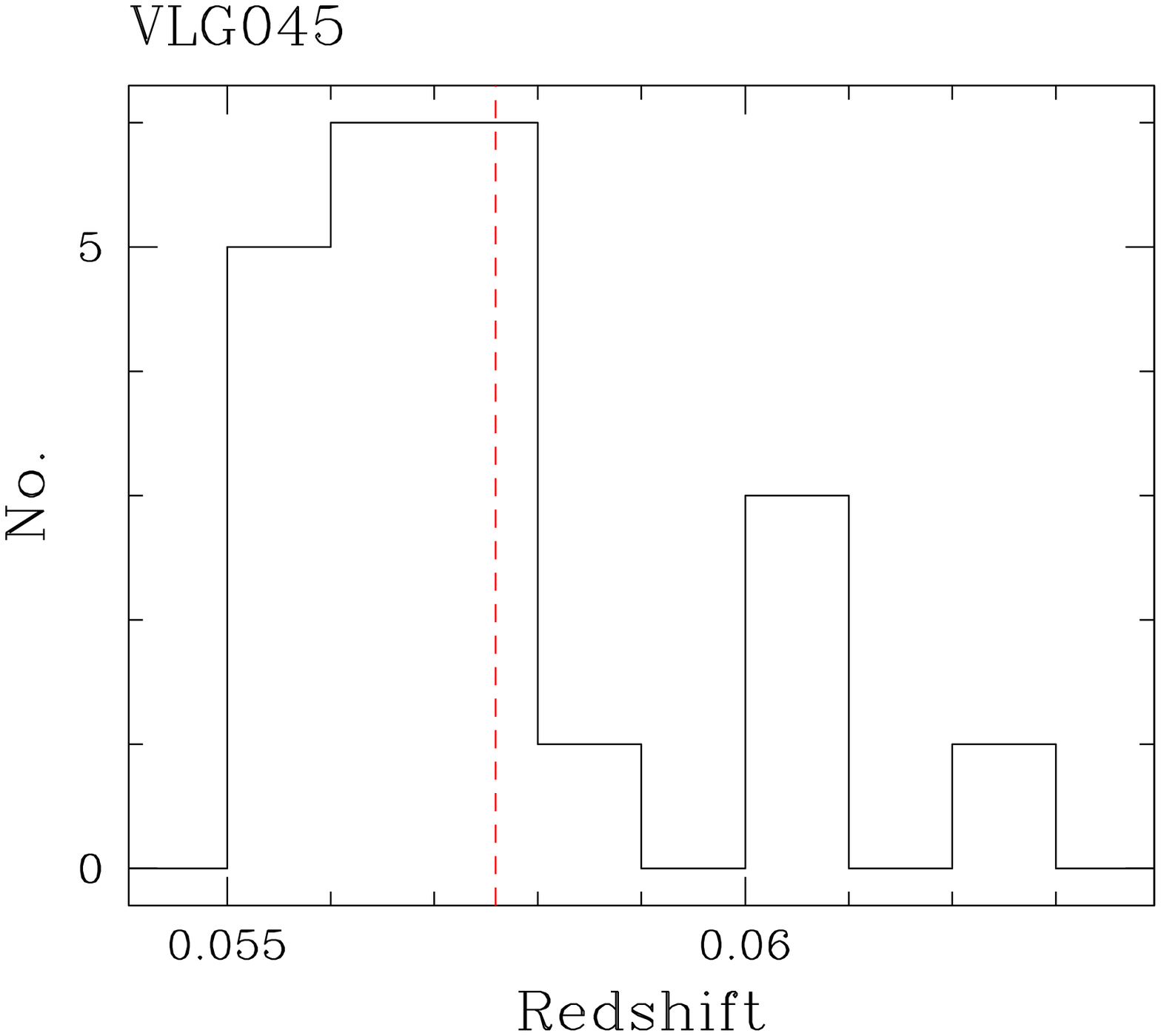} \hskip 0.1cm 
 \epsfxsize=4.1cm \epsfbox{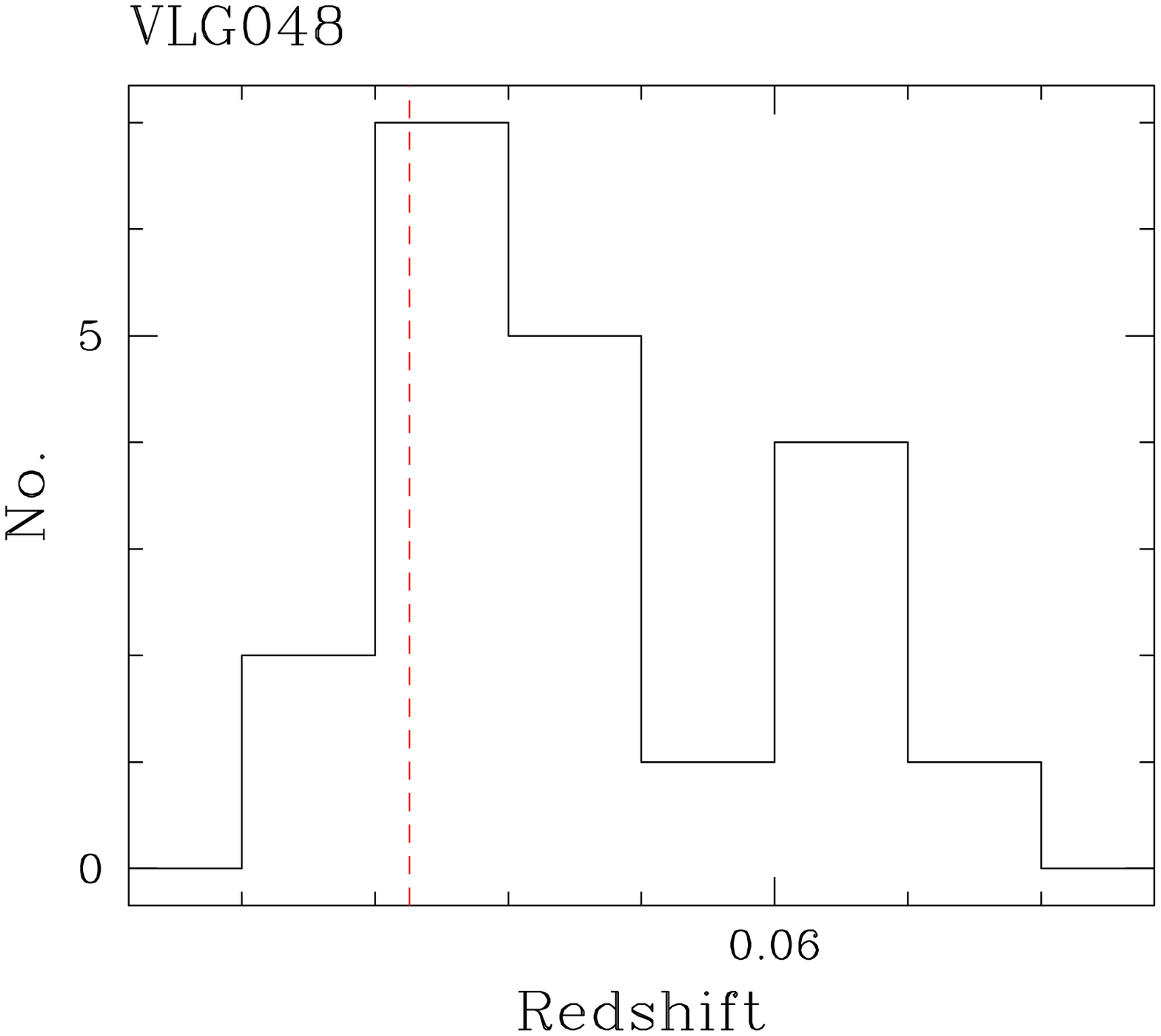}} \\
{\epsfxsize=4.1cm \epsfbox{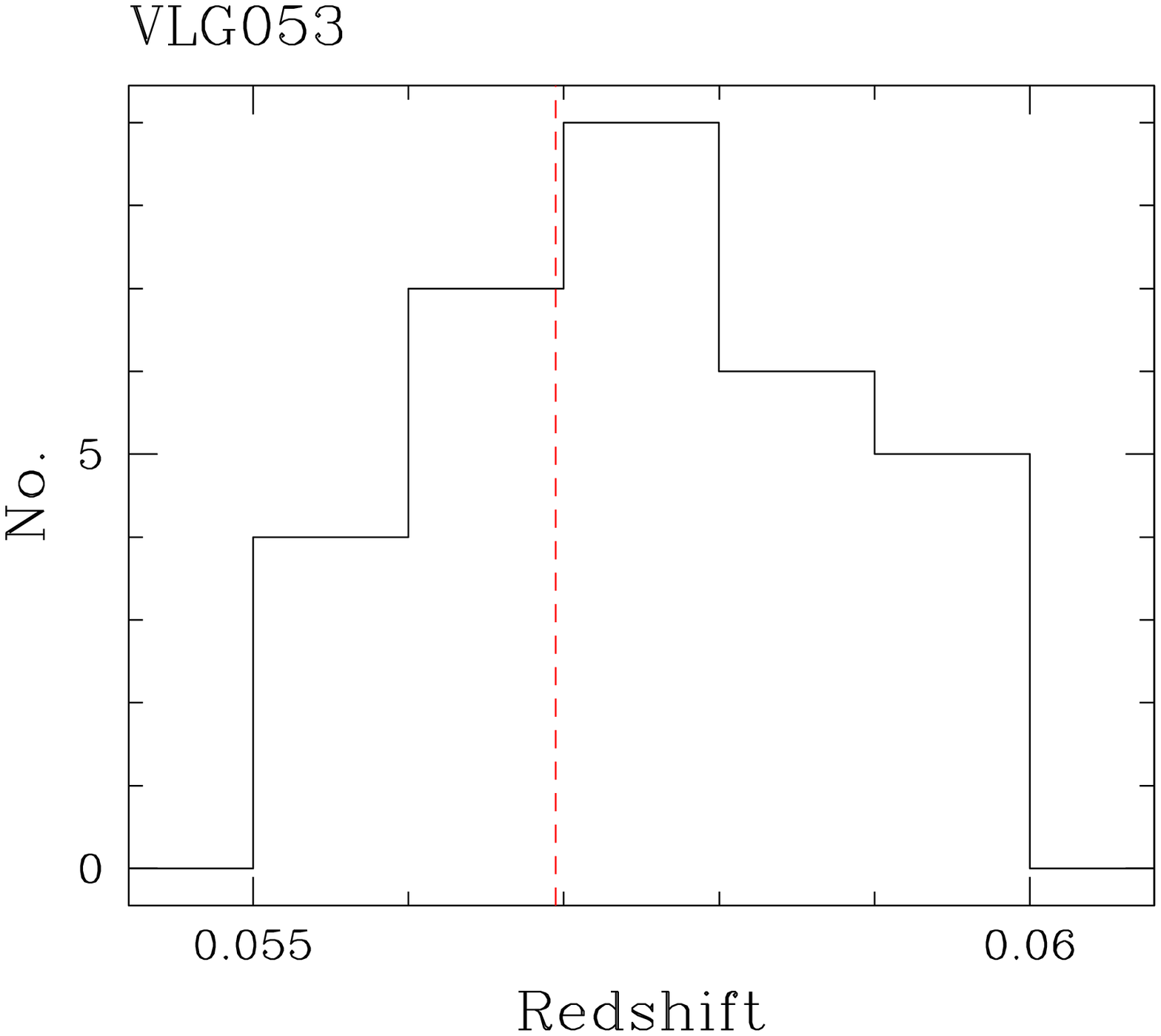} \hskip 0.1cm 
 \epsfxsize=4.1cm \epsfbox{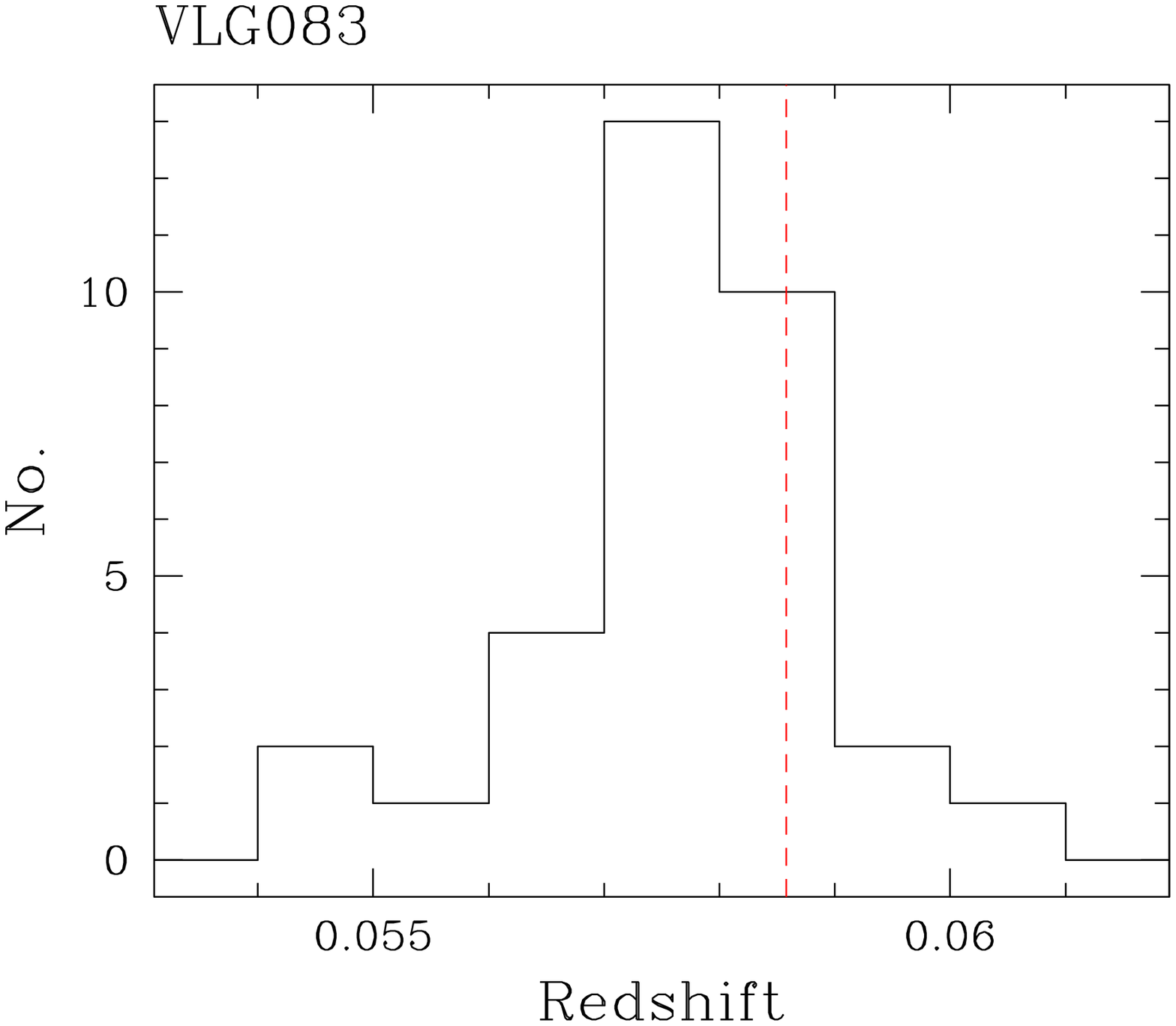} \hskip 0.1cm 
 \epsfxsize=4.1cm \epsfbox{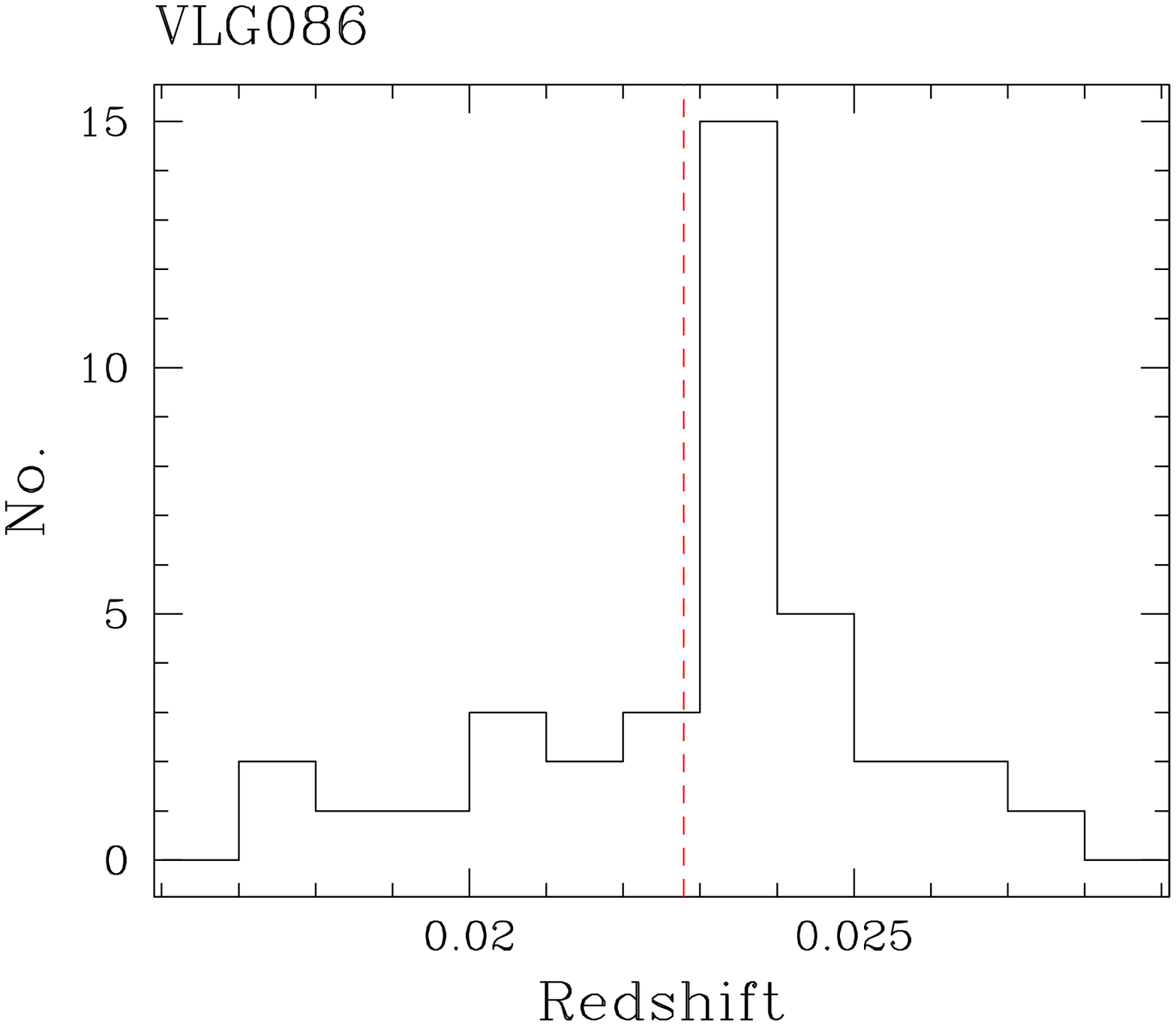}} \\ 
{\epsfxsize=4.1cm \epsfbox{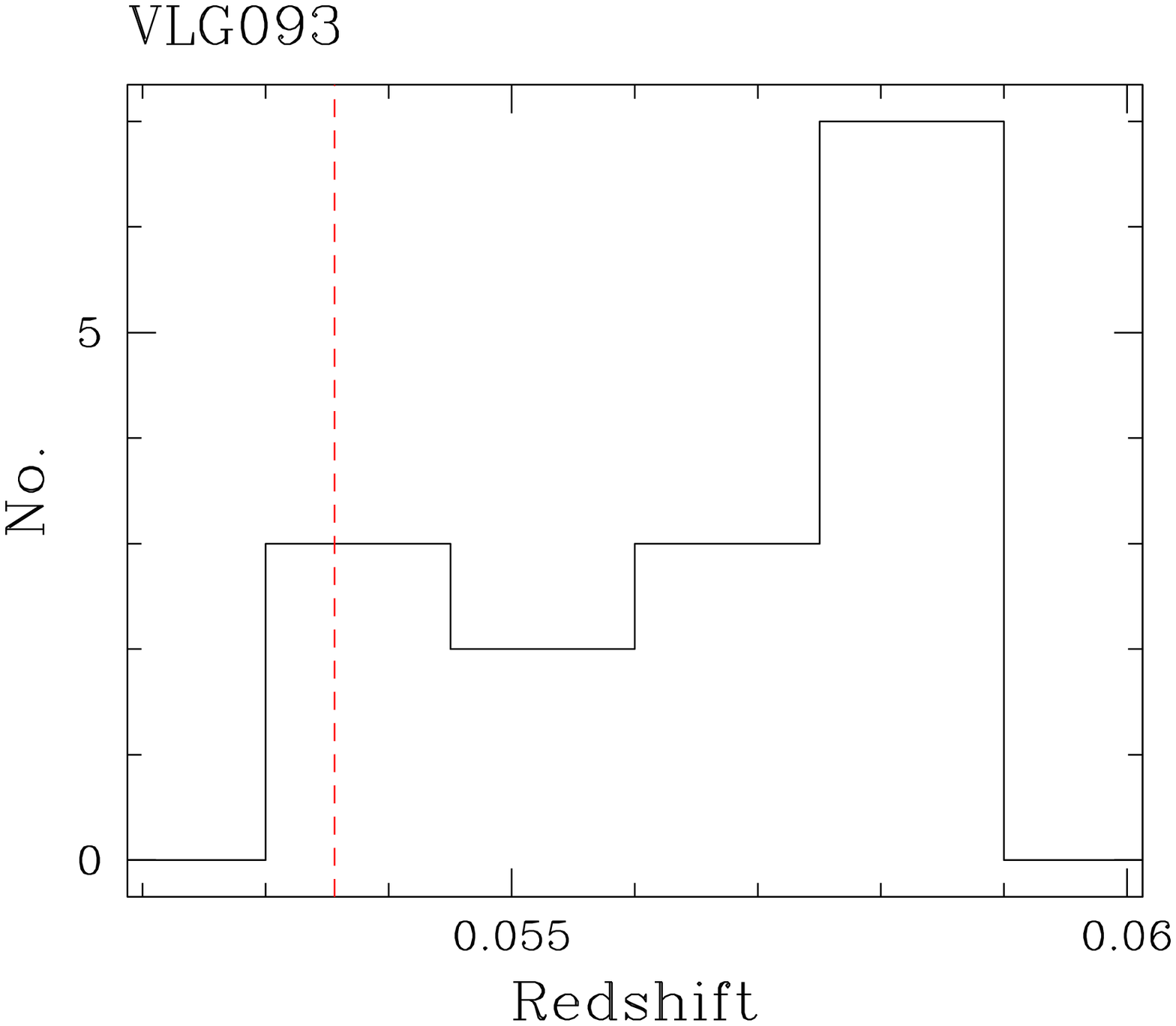} \hskip 0.1cm 
 \epsfxsize=4.1cm \epsfbox{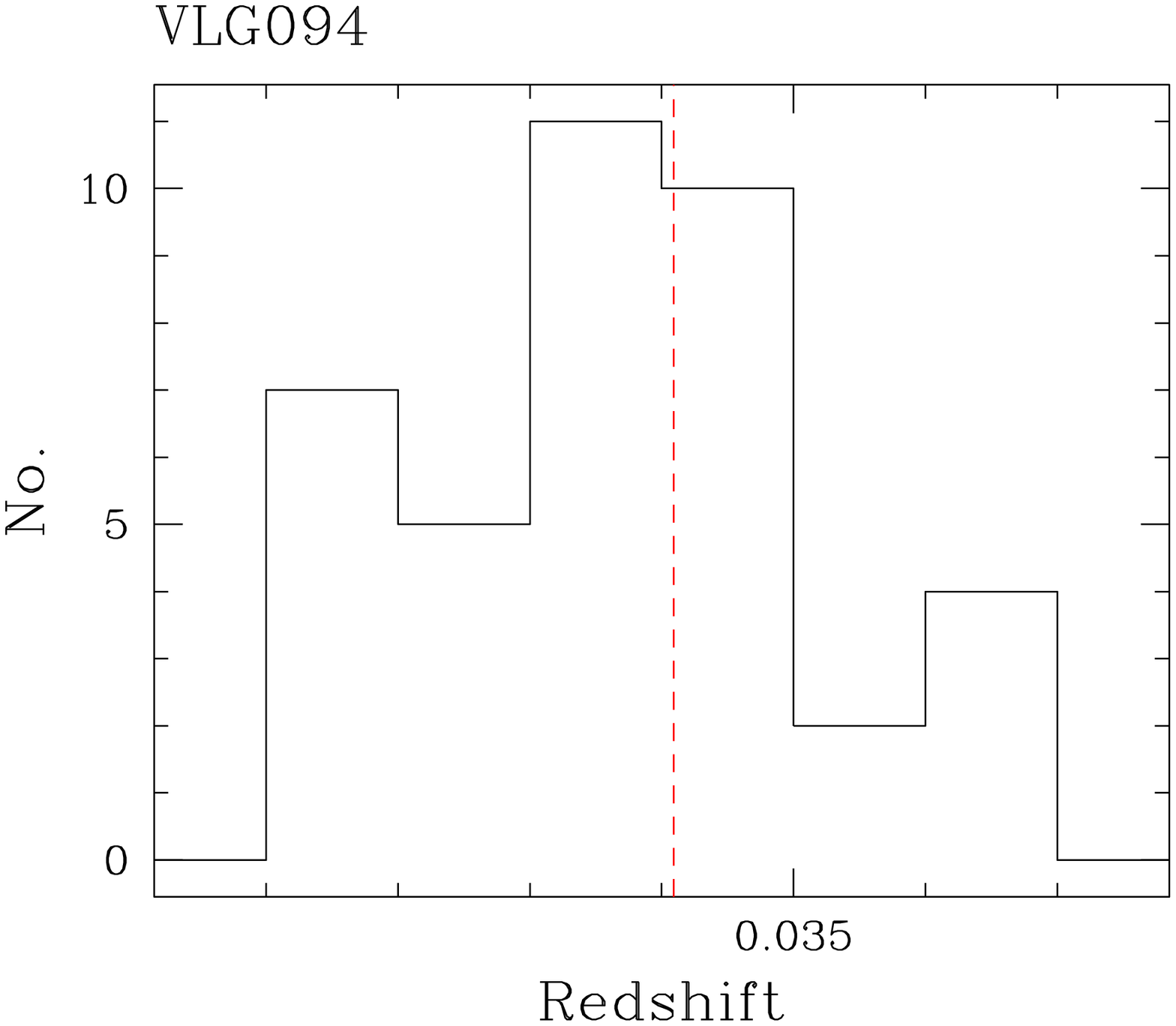} \hskip 0.1cm 
 \epsfxsize=4.1cm \epsfbox{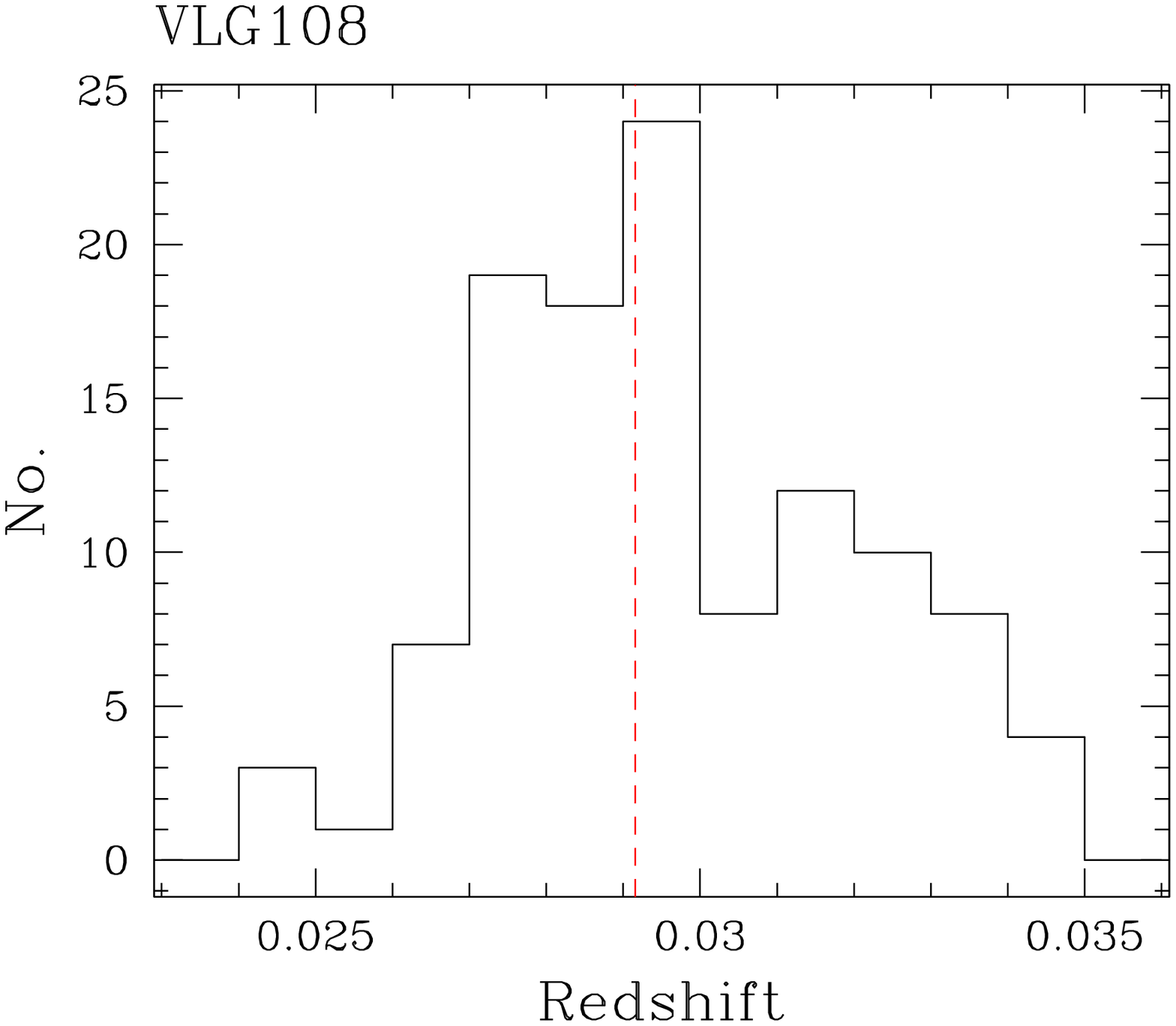}} \\
{\epsfxsize=4.1cm \epsfbox{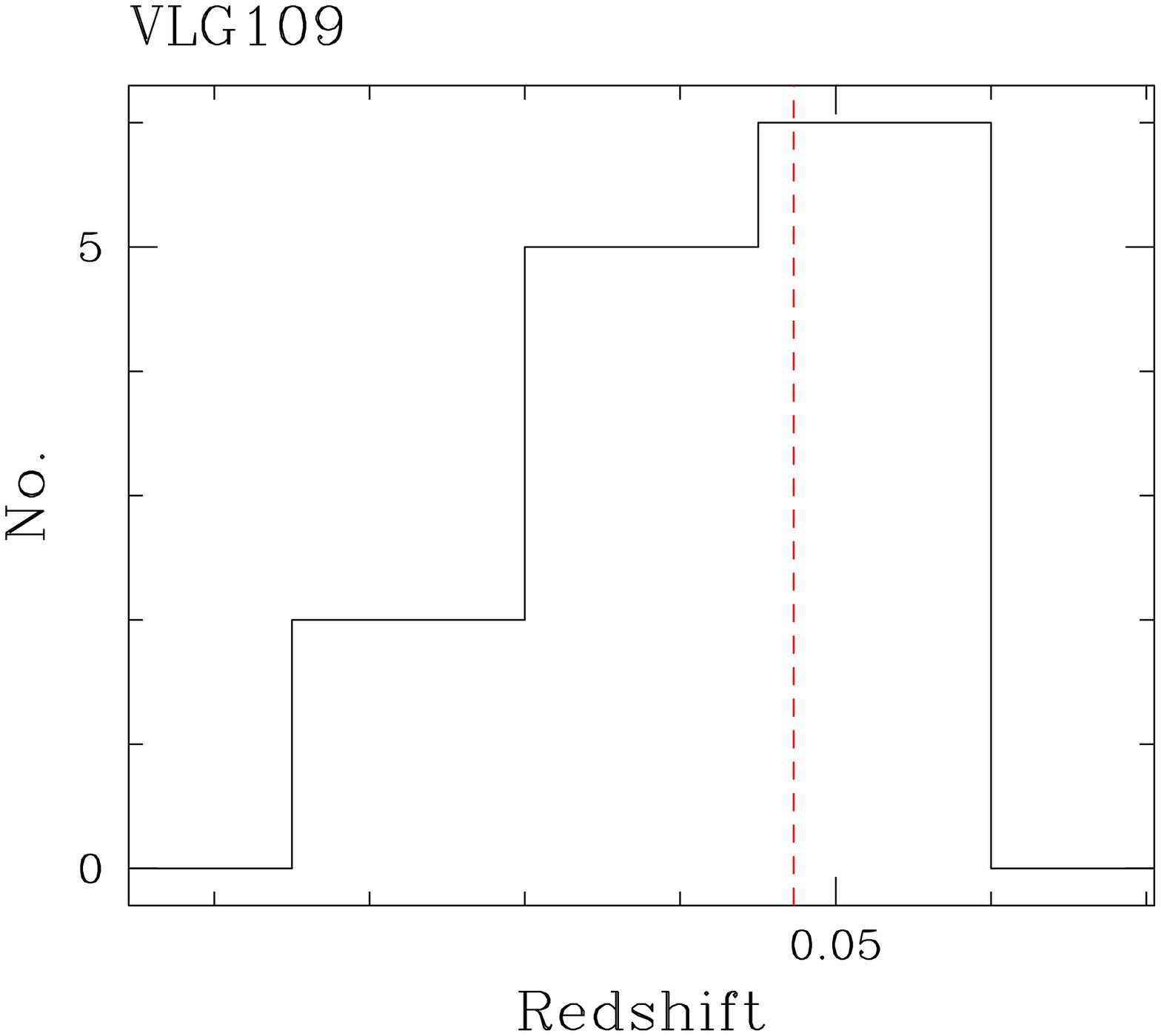}}
\end{figure*} 

In figure \ref{fig:hist} we show the velocity histograms of the systems with 
at least 10 measured redshifts. The redshift of the VLG is indicated by a
dashed line.

From table \ref{tab:systems}, and also taking into account the 4 VLGs
not included in this selection, it appears that all VLGs 
have companions. This is not so surprising, given the known 
correlation between galaxies, and confirms the idea that there are no truly
``isolated'' galaxies. Moreover, 
it is quite striking that in the literature 6 out of the 18 SSRS2 VLGs 
were not associated to any system, while other 3 were classified simply
as binaries: for example, VLG048 belongs to
 a system for which we could collect 15 new redshifts from the 2dFGRS.
Such a high number of nearby galaxy systems which were not previously
identified neither in the optical nor in the X--ray reminds us how poorly 
known the low--mass end of galaxy systems is.
Therefore it is interesting to examine in more 
detail the nature and environment in which VLGs are found.
In the following, we briefly describe the main properties of the selected
VLGs and their systems.

\begin{itemize}

\item{\bf VLG014} is an Sbc spiral in a known triplet, 
which appears to be in a larger system: 
including the information from the 2dFGRS, we 
have a total of 14 galaxies with redshift.
The 3--$\sigma$ clipping algorithm does not exclude any of these
galaxies, but 5 of them are apparently in foreground and 
background, contributing to the formally large velocity dispersion of the 
system, $\sim 750$ km/s; when considering only the central
9 galaxies (which include the VLG) concentrated at $z \sim 0.04$,
the velocity dispersion is only $\sim 120$ km/s. 

\item{\bf VLG022} is an S0 in a known compact group, SCG55, 
identified with an automated algorithm by Prandoni et al. (1997).
Five galaxies are within 1.5 \h ~in projected distance and 1500 km/s in
velocity from the VLG, but two are excluded by the 3$\sigma$ clipping.

\item{\bf VLG031}, a D galaxy, is the dominant galaxy of A151.
In our selection there is only another VLG (VLG108) 
associated to a rich Abell cluster. The VLG redshift is at the center of
the velocity distribution, which is regular and 
consistent with a Gaussian.
For this system we measure a velocity dispersion $\sigma \sim 720$ km/s.

\item{\bf VLG040} is a spiral in a system with a relatively large velocity 
dispersion (more than $700$ km/s); however,
the velocity histogram shows that this value is probably affected by
foreground galaxies. The VLG is in the concentration at $z \sim 0.043$.

\item{\bf VLG043} is an Sb galaxy previously not associated to a known 
system; now we have identified 
a group with 15 galaxies having measured redshifts, and a 
velocity dispersion of $\sim 470$ km/s. The velocity histogram is not
symmetric, with two galaxies which could be in the foreground, having
a velocity difference of about 1100 km/s with respect to the cluster mean.

\item{\bf VLG045} is a spiral classified in our original 
catalog as an ``isolated'' galaxy, 
as no known system was found in the literature.
Within the search radius we have found other 20 2dFGRS galaxies, and
only one of them was rejected by the 3-$\sigma$ clipping algorithm.
The velocity histogram presents a well defined peak, with
4 galaxies in the background.
The velocity dispersion of the system is $\sim 460$ km/s.
All these galaxies are much fainter than the VLG, which has an apparent 
magnitude of $14.96$: the second brightest galaxy has $b_J = 16.66$, 
while the others are fainter than $b_J=17$. Notice however that these fainter 
galaxies have still a significant intrinsic luminosity:
the second member has an absolute magnitude brighter than $M^*$.

\item{\bf VLG048} is an Sbc galaxy member of a known binary system:
however, with the 2dFGRS data we have 20 galaxy 
redshifts for this system.
The velocity histogram shows a main peak including the VLG, and a secondary 
peak in the background.

\item{\bf VLG053} is an S0 galaxy previously not associated to a 
known system, but looking at
the Digital Sky Survey image we had identified a few other bright galaxies. 
We can now confirm the reality of a group, with other 32 2dFGRS galaxies: 
its velocity histogram is consistent with a Gaussian 
distribution, with a velocity dispersion of $\sim 360$ km/s.

\item{\bf VLG061} is a large spiral galaxy, and also in this case, while
it was not associated to any known group or cluster, we had identified
many fainter galaxies in the field; 6 of them were selected and
observed at OHP, and their redshifts were found to be comparable to that
of the VLG. The velocity dispersion of the system is $\sim 300$ km/s. 
Galaxy $1$ (see table 1) has
a bright HII region outside the nucleus; in its spectrum we have 
detected $H_\beta$ in emission, 
the two [OIII] lines at 4959 and 5007 \AA, $H_\alpha$ and [NII] lines
(while [OII]3727 is outside our spectral range).

\item{\bf VLG068} was already known to be in a group:
we could only measure the redshifts of the three main components. 
However, we show this system
as a further example of how different can be the environment of a VLG from 
that of a typical cluster. The main companions of the VLG are
two Markarian galaxies, already known in the literature.
MRK609 is a Seyfert 1.8, with prominent emission lines 
($H_\beta$, [OIII] and $H_\alpha$).

\item{\bf VLG069} is a peculiar spiral galaxy in interaction in 
a binary system.
Its redshift is quite precise, having been measured also in $H_\alpha$.
From the 2dFGRS we have obtained a few other redshifts, but
there is some problem with the identification of the binary.
At the position of the VLG, the 2dFGRS gives a galaxy with a redshift 
consistent with the SSRS2, but the magnitude is significantly underestimated.
We have kept the SSRS2 measure in the table. The 2dFGRS gives also the redshift
of a second nearby galaxy, with coordinate approximately corresponding to the
center of the binary system and a bright magnitude (14.09). It could 
correspond to the companion of the VLG, and we have included it in our table.
In conclusion, 8 galaxies 
are now included as members of the group after the 3--$\sigma$ clipping.
This system has a relatively low velocity dispersion (160 km/s).

\item{\bf VLG074} is in fact 
a pair of interacting galaxies (NGC1516 A \& B), also detected by IRAS.
The redshift was measured by Strauss et al. (1992), attributing the 
coordinates to the center of the pair. 
These coordinates and redshifts were included in the SSRS2 and in our VLG 
catalog. It is possible that the magnitude of the brightest member
was not correctly estimated, 
given the proximity of the two galaxies.
We have measured the redshifts of both, 
which differ by less than 100 km/s (see table 1). A third galaxy among
those we have observed belongs to 
the system, while the other, fainter galaxies in the field are in the 
background, with a concentration at $z \sim 0.121$.  

\item{\bf VLG083} is an elliptical galaxy within 0.5 arcmin
from the center of the poor Abell cluster S0983.
In this field we have found other 30 2dFGRS galaxies:
the velocity dispersion of the system is $\sim 320$ km/s.

\item{\bf VLG086} is a Seyfert galaxy in a Hickson Compact Group (HCG91).
The VLG is in interaction with a nearby SB0.
This group is quite rich: 37 galaxies have measured 
redshifts and none of them is rejected by the 3--$\sigma$ clipping algorithm,
giving a quite large velocity dispersion of $\sim 670$ km/s. However, there is
a well defined central peak at $z \sim 0.023$, which corresponds to the mean 
redshift of the system and also to the redshift of the VLG. This should
probably be considered as the main group.
Note that we have excluded from our analysis a 2dFGRS galaxy, identified in the
NASA Extragalactic Database as 2dFGRS S175Z138, with coordinates
($\alpha = 22^h 09^m 07.45^s$, $\delta = -27^o 48^\prime 22.8^{''}$) ,
near to --but not coincident with-- the VLG center. It
might be identified with the VLG or alternatively with the SB0 interacting 
with the VLG.

\item{\bf VLG093} is an Sb in a field for which we have now
15 redshifts. The 3--$\sigma$ clipping does not exclude any of
these galaxies; however, the histogram is asymmetric, and the VLG,
not present in the 2dFGRS catalogue, is in the first velocity bin, while
the peak is in the last bin (the velocity difference is $\sim 1500$ km/s).
Therefore we can conclude that there is a group in the field, but it is not 
clear if it is really associated to the VLG.

\item{\bf VLG094} is a peculiar spiral in a poor cluster listed in the 
Edinburgh/Durham Cluster Catalog (EDCC155; Lumsden et al. 1992). The redshift 
of the VLG is comparable to the average redshift and is near the velocity peak.
The velocity dispersion, $\sigma \sim 410$ km/s, is quite typical of a rich 
group.
 
\item{\bf VLG108} is the cD galaxy in the Abell cluster A4038.
The usually quoted velocity dispersion of this cluster 
is larger ($\sim 882$ km/s in Struble \& Rood 1999) than our estimate of 
$\sim 660$ km/s.
However, as shown by Fadda et al. (1996), this 
apparently regular cluster has two different peaks in velocity (the 
two peaks can also be seen in our velocity histogram, see figure  
\ref{fig:hist}). 
Our velocity dispersion is in excellent agreement with
the X--ray temperature,  $kT = 3.31$ keV 
(Finoguenov et al. 2001); for example, using the Lubin \& Bahcall 
empirical relation (1993), $\sigma = 10^{2.52} (kT)^{0.6}$, 
we would expect $\sigma \sim 679$ km/s.

\item{\bf VLG109} is an elliptical VLG in the poor cluster S1155, 
which has a velocity dispersion comparable to S0983 (see VLG083), 
$\sim 300$ km/s.

\item{\bf VLG 0716+5323}, which we selected from the CfA catalog, is 
another example of an early--type VLG. It is in a Zwicky cluster (Z1261), 
with an extended X--ray emission centered at 0.5 arcmin from the VLG. 
This cluster, detected by the Einstein satellite, is also included in the 
ROSAT Brightest Cluster Sample (Ebeling et al. 1998), with a luminosity 
$L_X = 0.81 \times 10^{44}$ erg/s.
For this cluster there was no previous measured redshift except for the VLG 
(Gregory \& Burns 1982). At OHP we could measure the redshifts of 9 galaxies,
VLG included.
The resulting velocity dispersion of the system, $\sigma=609$ km/s, 
is typical of a poor cluster.
The agreement of the velocity dispersion with the cluster temperature, 
$kT = 2.8$ keV (Ebeling et al. 1998), is excellent: from the 
Lubin \& Bahcall (1993) fit we would expect $\sigma = 614$ km/s.

\end{itemize}

\section{The relation between VLGs and their environment}

In paper I we have shown that the correlation
function of VLGs approaches that of clusters.
Various galaxies discussed above 
(VLG045, VLG048, VLG053) are within the Pisces--Cetus Supercluster, 
which appears as a prominent feature near the limiting depth of the
SSRS2 volume--limited sample of VLGs
(the presence of this structure might explain the excess of  the VLG
correlation amplitude measured in the SSRS2 with respect to the 2dFGRS).
The Pisces--Cetus supercluster is well traced by Abell and ACO clusters 
(Tully 1986), but none of our VLGs is a member of those clusters.
The VLG correlation function is large because 
VLGs trace large--scale fluctuations just as clusters do, not because they 
are in rich clusters.

\begin{figure}
\caption[]{Normalized luminosity function of systems with an early type VLG 
(red squares) and a late type VLG (blue triangles), with poissonian 
$1 \sigma$ error bars.}
\label{fig:fdl}
\epsfxsize=12cm \epsfbox{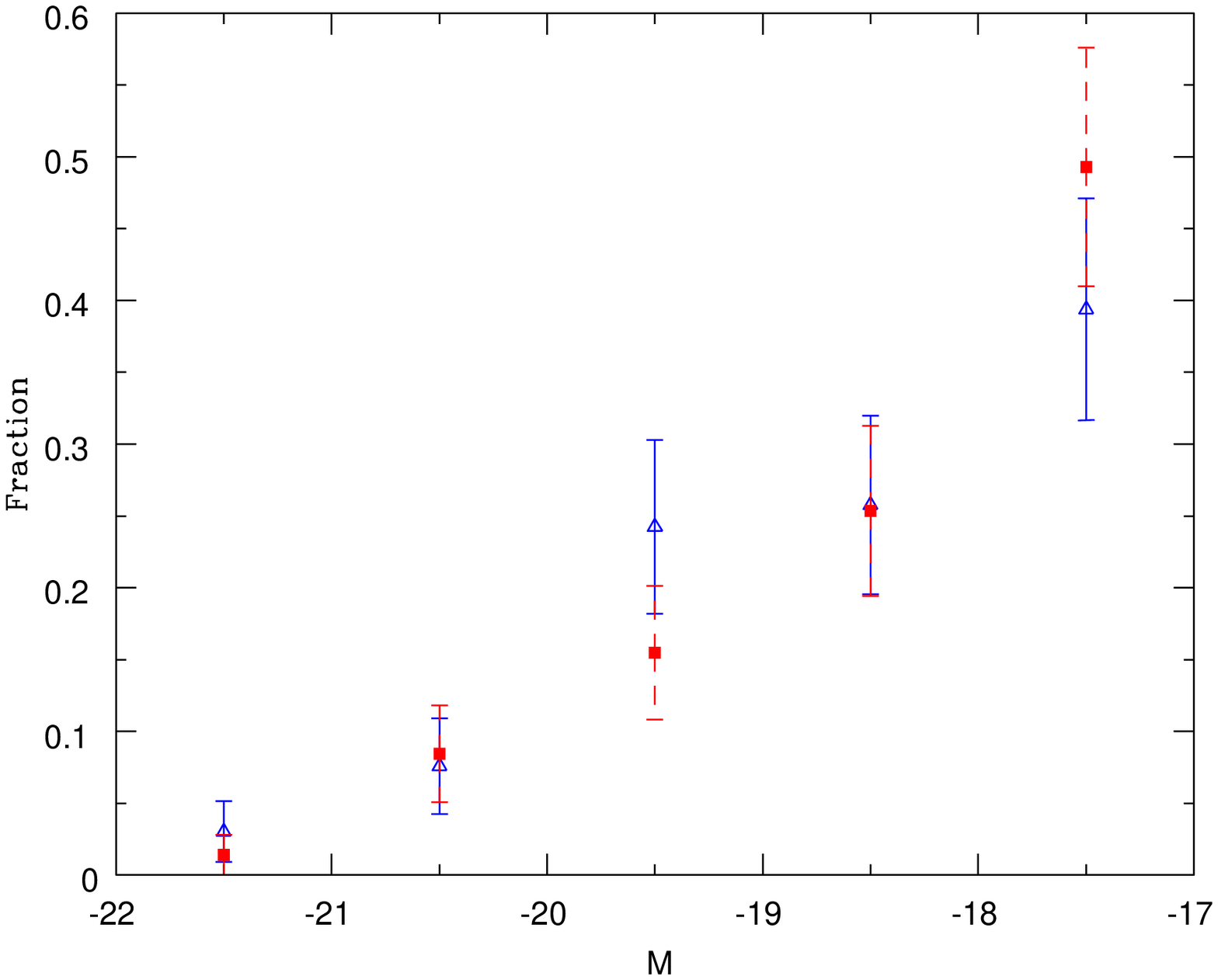}
\end{figure}

Another interesting issue is the luminosity function of the VLG systems.
In principle, merging should have played a major role in the formation
of the elliptical VLGs, and the luminosity function of the associated system
might be different from those dominated by a spiral VLG.
As a preliminary test, we have estimated the composite luminosity 
functions of early and late VLG dominated systems, excluding 
both rich clusters and systems with probable field contamination. In this rough
comparison, we assume that selection effects are the same for the two types
of systems. 
We have normalized numbers to the total number of galaxies in the two samples
(66 for spiral VLG and 71 for elliptical VLG systems).
In deriving absolute magnitudes, we applied the mean 
{\em K + e} correction formula
$K = 0.03 z /(0.01+z^4)$ adopted by Norberg et al. (2001); at the distances 
of SSRS2 VLGs ($\sim 0.05$), it is consistent with the correction adopted 
for the SSRS2 (da Costa et al. 1994; $K = 3z$). As apparent from figure
\ref{fig:fdl}, taking into account the small
number of objects in our samples we cannot find significant differences between
the two luminosity functions to $M=-17$, the absolute limiting magnitude at 
which galaxies around SSRS2 VLGs could still be detected in the 2dFGRS.

There is however a difference when looking at the morphology
of the central VLG and the velocity dispersion of the associated system.
It is not surprising that
the VLGs which are at the center of clusters are giant ellipticals; 
these clusters have $\sigma \sim 600$ km/s and
an associated X--ray emission at a temperature consistent with their
velocity dispersion. 
Other 3 systems including an early type VLG
(one S0 and two ellipticals) 
have velocity dispersions in the range 300--350 km/s,
and might be considered the low--mass end of galaxy clusters.
The remaining S0 in the sample is in a compact group with a small velocity 
dispersion ($\sim 140$ km/s).

The systems dominated by a spiral VLG have a large spread
in velocity dispersion: but as we have seen from the velocity histograms,
those with the largest 
velocity dispersions ($\sim 700$ km/s), VLG014, VLG040 and VLG086,
are probably affected by field contamination.
Moreover, for these systems no 
extended X--ray emission is reported in the on--line databases.

Within the limits of the small statistics and lack of completeness,
we can conclude that VLGs are in a qualitative agreement with the 
morphology--density relation: in clusters and rich groups we find 
only early type VLGs, while among systems originally classified as 
binaries or triplets, and those with the lowest velocity
dispersions (with the exception of the S0 VLG022) the VLG is a late--type.

The nearby systems we have observed at OHP have lower velocity dispersions, 
and have probably also a lower richness but the limiting apparent magnitude 
is also brighter for these systems.
The system more similar to our own Local Group is the one associated to
VLG069. It has 10 galaxies with measured redshifts and $M_B \le -15.5$: 
even if this is still not a complete sample, the number is comparable to
the Local Group, where we find 10 galaxies with $M_V \le -15.5$.
VLG069 is the main member of a binary system, as Andromeda and the 
Milky Way; moreover, the VLG069 system has also a relatively low velocity 
dispersion, $\sigma \sim 160$ km/s, still somewhat higher than 
the velocity dispersion of the Local Group,
$\sigma=61 \pm 8$ km/s (van den Bergh 1999, 2000), which is indeed lower than 
the observed range of our VLG systems. However, our observational errors
would not allow us to measure accurately such a low velocity dispersion.

Our optical classification of VLG systems can be compared to the
X--ray based classification of galaxy groups by Zabludoff \& Mulchaey 
(1998), who distinguish groups with a bright, central
elliptical galaxy and smooth X--ray emission from the hot IGM,
and groups without X--ray emission, a few bright late--type galaxies
with fainter members, like our Local Group.
Our systems with an early--type VLG have generally
velocity dispersion  and richness comparable to the values found by ZM98 for 
their 9 poor groups with diffuse X--ray emission. They also find that the 3
groups without X--ray emission have a lower number of members,
which seems also consistent with what we find. Zabludoff (1999) has suggested
the possibile existence of a third class of groups in a transition phase,
but it has still to be demonstrated that the differences between the two 
classes might be due to evolution instead of their different
formation processes.

\section{Discussion and conclusions}

The number density of galaxies is dominated by faint, small systems.
As an example, let us assume a Schechter luminosity function
with $M^* _{b_J}=-19.6$, $\alpha=-1.22$ and $\phi^*=0.02$ (Zucca et al. 1997).
The fraction of galaxies brighter than $L^*$ is in fact less than
2\% of all galaxies with $M_{b_J} \leq -12.5$.
The brightest galaxies with $M_{b_J} \leq -21$ are only 3 out of 10000.
Among these galaxies, we find M31 and probably the Milky Way
(see e.g. van den Berg 1999), which are therefore not typical galaxies, 
but very special systems.
Nevertheless, when looking at the luminosity (mass) density, 
the contribution of Very Luminous
Galaxies (VLGs) with $M_B \leq -21$ to the luminosity
density increases to 1.6\% and that of 
$M_*$ galaxies to nearly 30\%. VLGs are extremely interesting
from the point of view of galaxy formation and large--scale structure.
They are visible at large distances ($D_{max} \sim 170$\h
~at the limiting magnitude of the SSRS2 $m_B=15.5$) 
and their distribution is biased with
respect to galaxies of lower luminosity.

There is a common misconception, according to which
optically very luminous galaxies selected with a large 
correlation amplitude are early--type galaxies in clusters. 
In this work we have shown that, at least choosing galaxies in the blue
passband, this is not the case. 
We have presented our observations and 2dFGRS data concerning fields 
centered on SSRS2 VLGs: we have found clear evidence that VLGs are the 
brightest members of galaxy systems which can escape standard group finding
methods, except of course for the minority of early--type VLGs in
rich groups or clusters.
VLGs have clustering properties similar to clusters, but most of them are in
systems with a galaxy population comparable to loose groups, and some of them
are probably comparable to our Local Group.
The large correlation amplitude suggests that VLGs are in high density
regions; most of them, being spirals, 
cannot have accreted more than a few percent of their mass
through major merging episodes (T\'oth \& Ostriker 1992), so we have to 
suppose that
these systems already formed with a large, central massive galaxy and
low mass companions. On the other hand,  the merging
of two late--type VLGs could evolve into an early type system, analogously
to what was suggested for poor groups by Zabludoff \& Mulchaey (1999).

Other recent works appear to confirm the general properties of VLGs
which we have found from the analysis of the SSRS2.
Giuricin et al. (2001) have analysed the Nearby Optical Galaxy sample,
finding a similar trend for the luminosity segregation, and
that only 10\% of their VLGs reside in clusters; they also find that, while
the fraction of very luminous early--type galaxies is larger than 
the corresponding fraction for the total sample,
it is still only $29$\%, less than than the fraction of Scd galaxies (39\%).
Moreover, from the analysis of the 2dFGRS Norberg et al. (2002) confirm that 
``luminosity, and not type, is the dominant factor in determining how the 
clustering strength of the whole galaxy population varies with luminosity''.

The amount of mass associated to VLG systems is still an open question. 
For example, the lower correlation amplitude for VLGs found in the 
2dFGRS would indicate that VLGs are associated to dark halos less massive
than typical halo clusters.
The increase in local overdensity of galaxies around VLGs should also be better
determined.
In a recent paper Hogg et al. (2003) analyse the Sloan Digital Sky Survey 
and find the intriguing results that blue luminous galaxies with 
$L < 3 L^*$ are not apparently found in  overdensities, but VLGs have 
even larger luminosities.

Therefore only further and deeper observations devoted to the detailed study 
of VLGs, determining the luminosity function of 
these systems and the dynamics of satellites around the VLGs, together with 
observations in redder passbands (more representative of the mass of the 
systems) will shed more light on the properties of VLGs and their 
environment, and their implications for galaxy formation and evolution.

\acknowledgements
This work has been partially supported by the Italian Space Agency grants
ASI-I-R-105-00 and ASI-I-R-037-01, and by the Italian Ministery (MIUR)
grant COFIN2001 ``Clusters and groups of galaxies: the interplay between
dark and baryonic matter". We thank the referee, Florence Durret,
for her careful reading of the manuscript and useful comments.

\appendix

\section{The discrepant redshift of Arp 127}

\begin{figure}
\label{fig:arp}
\caption[]{Arp127: finding chart (the scale is $8 \times 8$ square arcmin).}
{\epsfxsize=8.0cm \epsfbox{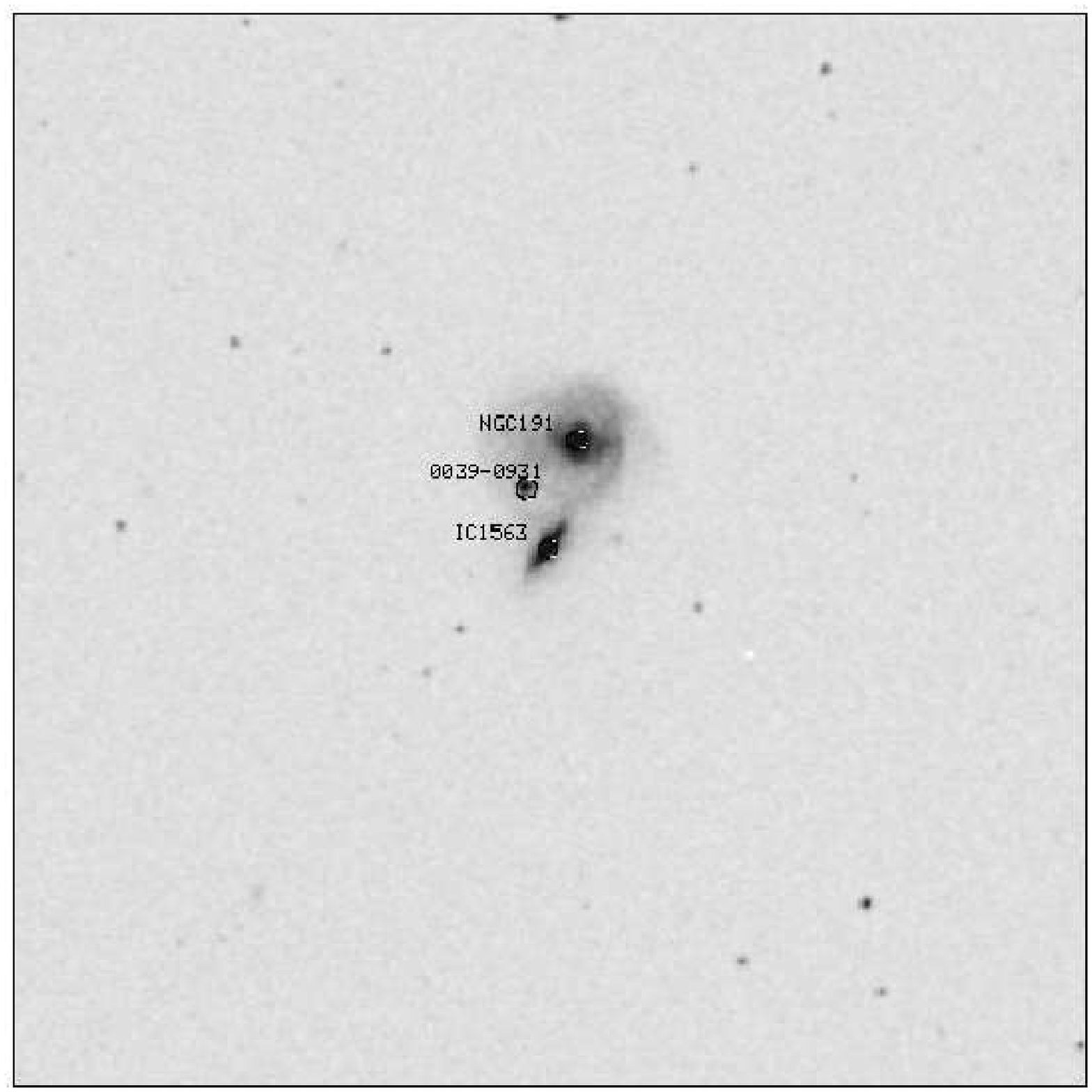}}
\end{figure}

The Arp 127 pair (Arp 1966) is made by NGC0191, a spiral classified as 
SAB(rs)c: pec, and IC1563, an S0 pec sp. A third, more compact object
is seen among the two galaxies (see \ref{fig:arp}). According to 
the literature, the redshifts of NGC0191 and
IC1563 are respectively $v=5065 \pm 141$ km/s and $v=13652 \pm 141$ km/s 
(Huchra et al. 1993), a surprising difference given the apparent signs of 
interactions; for this reason IC1563 was not included in our final catalogue, 
even if according to the quoted redshift and the 
apparent magnitude  ($m = 14.74$) it should be considered a VLG. 

Our measurements for NGC0191 and IC1563 are reported in table A.1, and show 
that there is no discrepancy:
the binary system Arp 127 is at $z \sim 6150$ km/s, and the two galaxies have
a velocity difference of only $\sim 60$ km/s --, i.e. they have the same
velocity taking into account the errors.
These two galaxies are clearly interacting, as shown by the tidal distortion
in NGC0191.

We suggest that
the redshift of $13652$~km/s should be attributed to the round object 
between the two galaxies. 
In fact Beers et al. (1991) report that value for the redshift giving the
coordinates of the round object, but identify it as IC1563, while Huchra 
et al. (1993) give the approximate coordinates of the Arp 127 system for both
NGC0191 and IC1563.
The velocity of the round object is in the lower part of the velocity range
of the A85 galaxy cluster (see Durret et al. 1998) 
and at an angular distance of 44 arcmin from its center, and it might be a 
galaxy member of the cluster.

The new redshift for IC1563 means that this galaxy has $M=-19.2$,
i.e. it is a typical $M^*$ galaxy and not a VLG.
In table \ref{tab:arp127} 
 we report also the photometric observations of Reshetnikov \& Combes (1996).  

\begin{table*}[ht]
 \caption[]{Heliocentric Redshifts of Galaxies in Arp 127}
 \label{tab:arp127}
 \begin{flushleft}
 \begin{tabular}{rrrrrrrr}
  \hline
  \hline
 Iden.  &   RA (2000) & DEC (2000) &  $V_h$ (km/s) & Error & $B_t$ Notes \\
 \hline
%--------------------------------------------------------------- start data
 NGC0191   & 00 38 59.3 & -09 00 09 & ~6076 & ~32 & 12.5  [13.68] &     \\
 IC01563   & 00 39 00.2 & -09 00 53 & ~6138 & ~39 & 14.74 [14.39] &        \\
 0039-0931 & 00 39 02.0 & -09 00 31 & 13652 & 141 &       &                \\
%--------------------------------------------------------------- end data
\hline
\end{tabular}
\end{flushleft}
\end{table*}

\section{2dFGRS data}

In the following tables we list the 2dFGRS galaxies which we have selected as 
members of VLG systems. We give in column (1) our identification number, 
in columns (2) and (3) respectively right ascension and declination, in column
(4) the $b_J$ magnitude, and in column (5) the redshift. 

We have listed
the 2dFGRS data on the VLG when available, otherwise we have reported the
SSRS2 data.

We could compare magnitudes and redshifts of the 2dFGRS and SSRS2 for
5 VLGs with 2dFGRS data (VLG014, VLG022, VLG043, VLG045, VLG048, VLG109):
we find an average velocity difference 
$<V (2dF) - V (SSRS2)> = 118$ km/s with an rms of $155$ km/s and
an average magnitude difference $<b_J (2dF) - m_B (SSRS2)> =
 0.20$, with an rms of 0.46. 
The magnitude difference is consistent with the $\sim 0.2$ zero--point shift
expected between the blue magnitudes of the SSRS2 
(Alonso et al. 1993, 1994; da Costa et al. 1994) 
and the APM $b_J$ magnitudes on which the 2dFGRS is based
(Maddox et al. 1990a, 1990b, 1996). Note that for bright galaxies, 
magnitudes are
not very precise: the APM $b_J$ magnitudes have a precision of
0.2 in the range $17 - 19.5$ but are significantly affected by saturation at
magnitudes brighter than $b_J = 16$ (Norberg et al. 2002).
 
{\bf Tables available at http://www.bo.astro.it/$\sim$cappi/preprints.html}

\end{document}